\documentclass[superscriptaddress,amsmath,amssymb,aps,prb,citeautoscript,10pt]{revtex4-2}
\usepackage[utf8]{inputenc}% Language package
\usepackage{graphicx}% Include figure files
\usepackage{bm}% Bold math
\usepackage[colorlinks,allcolors=blue,breaklinks]{hyperref}% Add hypertext capabilities
\graphicspath{{./pics/}}
\usepackage{color, colortbl}
\usepackage[table, svgnames, dvipsnames]{xcolor}
\usepackage{makecell, cellspace}
\definecolor{Gray}{gray}{0.9}
\usepackage{xcolor}
\newcommand{\revision}[1]{#1}
\edef\restoreparindent{\parindent=\the\parindent\relax}
\usepackage{parskip}
\restoreparindent
\usepackage{filecontents}

\usepackage{pdfpages}
\makeatletter
\AtBeginDocument{\let\LS@rot\@undefined}
\makeatother

\usepackage{titlesec}
\titleformat{\section}{\normalfont\fontsize{11}{11}\bfseries}{}{}{}[]
\titlespacing{\section}{0em}{9pt}{3pt}
\titleformat{\subsection}[runin]{\normalfont\fontsize{10}{10}\bfseries}{}{}{}[.]
\titlespacing{\subsection}{0em}{9pt}{4.4pt}
\usepackage{tikz}

\begin{document}
\setcitestyle{super}  % A nature-like citation style
\title{Inferring topological transitions in pattern-forming processes with self-supervised learning}

\author{Marcin Abram}
\email{mjarbam@usc.edu}
\affiliation{Department of Physics and Astronomy, University of Southern California, Los Angeles, CA 90089, USA}
\affiliation{Information Sciences Institute, University of Southern California, Marina del Rey, CA 90292, USA}

\author{Keith Burghardt}
\email{keithab@isi.edu}
\affiliation{Information Sciences Institute, University of Southern California, Marina del Rey, CA 90292, USA}

\author{Greg Ver Steeg}
\affiliation{Information Sciences Institute, University of Southern California, Marina del Rey, CA 90292, USA}

\author{Aram Galstyan}
\affiliation{Information Sciences Institute, University of Southern California, Marina del Rey, CA 90292, USA}

\author{Remi Dingreville}
\email{rdingre@sandia.gov}
\affiliation{Center for Integrated Nanotechnologies, Nanostructure Physics Department, Sandia National Laboratories, Albuquerque, NM 87185, USA}
%%%%%%%%%%%%%%%%%%%%%%%%%%%%%%%%%%%%%%%%%%%%%%%%%%%%%%%%%%%%%%%%%:~

\date{\today}

\begin{abstract}
The identification and classification of transitions in topological and microstructural regimes in pattern-forming processes \revision{are} critical for understanding and fabricating microstructurally precise novel materials in many application domains. Unfortunately, relevant microstructure transitions may depend on process parameters in subtle and complex ways that are not captured by the classic theory of phase transition. While supervised machine learning methods may be useful for identifying transition regimes, they need labels which require prior knowledge of order parameters or relevant structures \revision{describing these transitions}. Motivated by the universality principle for dynamical systems, we instead use a self-supervised approach to solve the inverse problem of predicting process parameters from observed microstructures using neural networks. This approach does not require \revision{predefined,} labeled data about \revision{the different classes of microstructural patterns or about} the target task of predicting microstructure transitions. We show that the difficulty of performing the \revision{inverse-problem} prediction task is related to the goal of discovering microstructure regimes, because \emph{qualitative changes in microstructural patterns} correspond to \emph{changes in uncertainty predictions} for our self-supervised problem. We demonstrate the value of our approach by automatically discovering transitions in microstructural regimes in two distinct pattern-forming processes: the spinodal decomposition of a two-phase mixture and the formation of concentration modulations of binary alloys during physical vapor deposition of thin films. This approach opens a promising path forward for discovering and understanding unseen or hard-to-\revision{discern} transition regimes, and ultimately for controlling complex pattern-forming processes.
\end{abstract}

\maketitle

%%%%%%%%%%%%%%%%%%%%%%%%%%%%%%%%%%%%%%%%%%%%%%%%%%%%%%%%%%%%%%%%%
%%%%%%%%%%%%%%%%%%%%%%%%%%%%%%%%%%%%%%%%%%%%%%%%%%%%%%%%%%%%%%%%%
%%
%% INTRODUCTION
%%
%%%%%%%%%%%%%%%%%%%%%%%%%%%%%%%%%%%%%%%%%%%%%%%%%%%%%%%%%%%%%%%%%
%%%%%%%%%%%%%%%%%%%%%%%%%%%%%%%%%%%%%%%%%%%%%%%%%%%%%%%%%%%%%%%%%
\section*{INTRODUCTION}

\par{
Identifying and characterizing transitions in pattern-forming processes \revision{are} key to our understanding and control of many systems in physical, chemical, material, and biological sciences.
\revision{These tasks} can be particularly challenging when the transition from one type of microstructural pattern to another is continuous or ambiguous as may be the case in dynamic and second-order phase transition processes.
The reason is the state of the material system may exhibit subtle changes in the microstructure when the process parameters controlling the formation of patterns vary.
For example, in cellular dynamics, cytoskeletal protein actin filaments undergo a continuous isotropic to nematic liquid crystalline phase transition when polymerized~\cite{viamontes2006isotropic}.
In chemical systems undergoing a precipitation reaction (Liesegang systems), one can control the transition from one precipitation pattern to another by regulating solid hydrogel and reactant concentrations~\cite{antal1999formation, toramaru2003experimental, shimizu2017liesegang, nabika2019pattern}.
When heated up, block copolymers thin films experience a smooth, thermally induced morphology transition from cylindrical to lamellar microdomains~\cite{sakurai1993morphology, castelletto2004morphologies}.
Similarly, the co-sputter deposition of immiscible elements results in a variety of self-organized microstructural patterns depending on processing conditions~\cite{Lu2012, herman2020data, powers2020microstructural, powers2021compositionally}.
}

\par{
The theory describing these types of transitions was first proposed by Landau~\cite{landau1937theorie} and later improved using renormalization group theory~\cite{muller1999variational}.
In such models, the determination of a pattern transition relies on 
(i) the identification of a local order parameter that sharply changes from one value to another (\textit{i.e.} the order parameter changes discontinuously as in first-order transitions) and
(ii) straightforward symmetry-breaking considerations of that order parameter.
However, in many cases, topological (structural) transitions~\cite{kosterlitz1973ordering, bagchi1996computer, bel2021geometrical,Stewart2020} are gradual and elusive, making it more difficult to identify the appropriate indicators of microstructural topology transitions.
Supervised learning techniques,~\cite{carrasquilla2017machine,wei2017identifying, li2018applications, casert2019interpretable, zhang2019machine} that use \revision{predetermined} labelling of distinct types of patterns for \revision{given} process parameters, are commonly used to classify and predict transitions.
However, generating these labels requires prior knowledge of the regimes in the process parameter space where the transition may occur.
This limits the scope of this class of methods primarily to already \revision{known} and at least partially \revision{and previously explored transition} cases.
In contrast, unsupervised learning techniques~\cite{wang2016discovering, hu2017discovering, Nieuwenburg2017, wetzel2017unsupervised, liu2018discriminative, yoshioka2018learning, rodriguez2019identifying, lee2022phase} do not require hand labelling or other time-consuming manual and potentially subjective interventions.
Many of these unsupervised approaches use dimensionality reduction and clustering to classify topological phases in latent space and detect the topological transitions.
}

\par{
Like unsupervised methods, self-supervised learning does not require \revision{predefined} labels for the target task.
Instead of directly solving a clustering problem, self-supervised learning focuses on solving an auxiliary prediction problem which \revision{is easy to measure and closely related and semantically connected to the target task}.
Often the auxiliary problem requires predicting structure of the training data instead of some external labels, hence the ``self-supervised'' name.
The promise of this method is that, while learning to solve the auxiliary tasks\revision{,} for which labels can be easy to generate and do not require human supervision\revision{,} we can learn the structure of the original problem for which the true labels are unknown.
\revision{In the present case, we aim at learning the structure of the data in pattern-forming processes and use this knowledge to identify the transitions between different classes of patterns for which we do not have labels.}
\revision{Self-learning approaches} can lead to models achieving comparable performance as when trained in a traditional supervised fashion, while reducing reliance on \revision{predefined} labeled data \cite{doersch2017multi, zbontar2021barlow}.
In computer vision for instance, learning to predict whether two images are modified versions of the same original image, allows to construct a high-quality latent representation of the data, that can subsequently be used for solving an object recognition task \cite{chen2020simple}.
}

%%%%%%%%%%%%%%%%%%%%%%%%%%%%%%%%%%%%%%%%%%%%%%%%%%%%%%%%%
%%% FIG 1: WORKFLOW
%%%%%%%%%%%%%%%%%%%%%%%%%%%%%%%%%%%%%%%%%%%%%%%%%%%%%%%%%
 \begin{figure*}
    \centering
    \includegraphics[width=\textwidth]{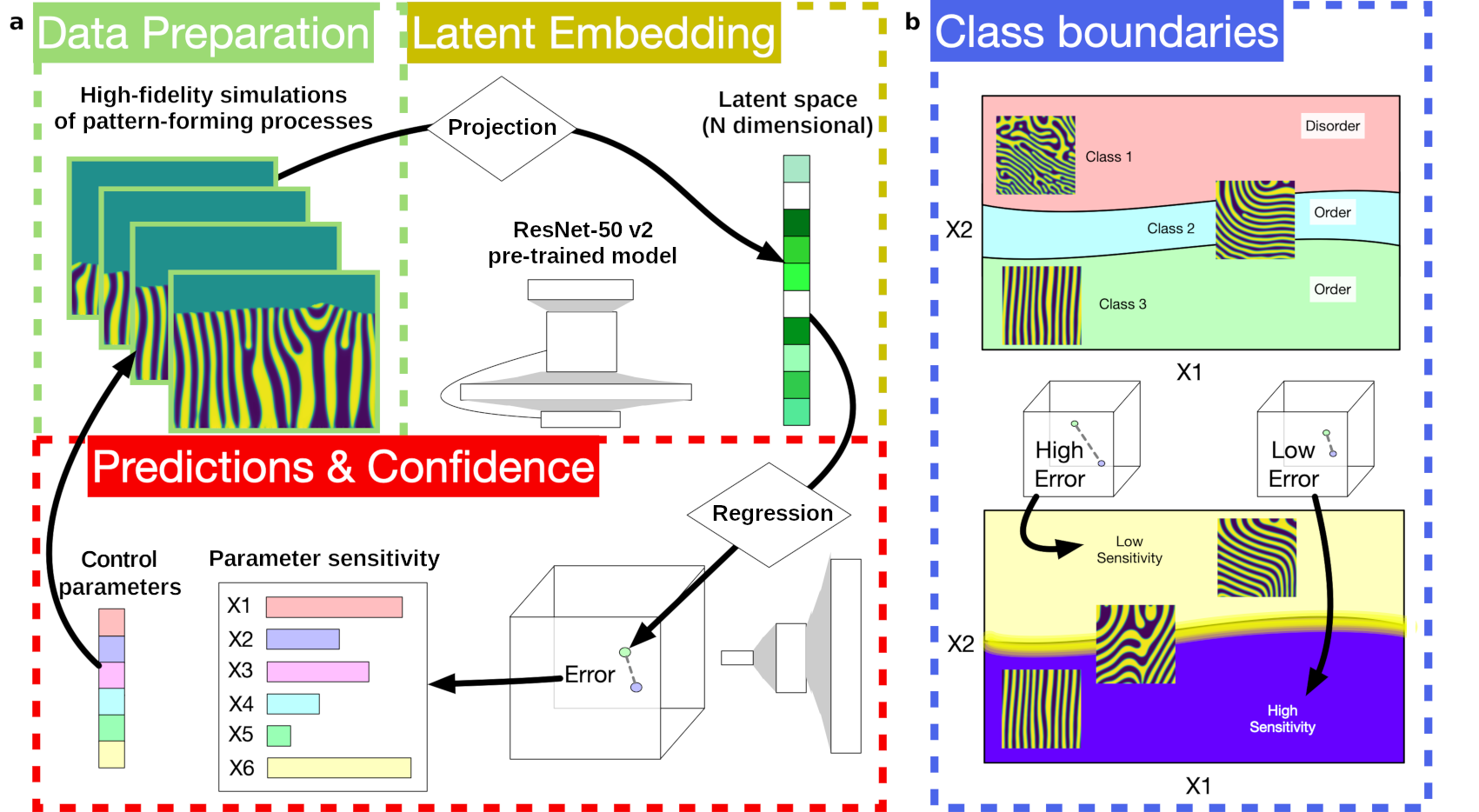}
    \caption{{\bf Workflow to identify transition regimes in pattern-forming processes via self-supervised learning}.
      {\bf a}~We~simulate the dynamical evolution of the physical system for a broad range of process parameters. Next, we project the final state of the microstructural pattern into a latent space (using a pre-trained ResNet-50 v2 \cite{He2016}). We regress on these latent dimensions to estimate the original process parameters.
      {\bf b}~To detect specific classes of microstructural patterns, we evaluate the model error by predicting the corresponding initial process parameters.
      By measuring the change in sensitivity of forming specific patterns changes for various input process parameters, we learn where the transition regime(s) might occur.
      }
    \label{fig:workflow}
 \end{figure*}
%%%%%%%%%%%%%%%%%%%%%%%%%%%%%%%%%%%%%%%%%%%%%%%%%%%%%%%%%
%%%%%%%%%%%%%%%%%%%%%%%%%%%%%%%%%%%%%%%%%%%%%%%%%%%%%%%%%

\par{
With all its advantages, it can be challenging to define an appropriate task to use for the self-supervised training.
In the context of the present work, our target task is the identification and characterization of microstructure transitions \revision{when the process parameters controlling the pattern formation vary}.
We propose that a relevant auxiliary task is to solve the inverse problem of predicting input process parameters from observed microstructures.
The theoretical motivation behind this inverse problem stems from the fact that in material systems that experience a topological transition (or critical point), the sensitivity of forming specific patterns varies depending on the input process parameter.
This sensitivity can sometimes increase, as it would be the case if \revision{a given} process parameter acts as an order parameter, \revision{uniquely characterizing a regime for which a specific pattern exists}.
Or the sensitivity can decrease, as it would be the case for systems that exhibit universality (\textit{i.e.} the closer the parameter is to its critical value, the less sensitively the order parameter depends on the dynamical details of the system).
Through our study, we found the inverse problem to be strongly related to our target task of identifying microstructure transitions.
Interestingly, the most relevant \revision{signature} for identifying microstructure regimes emerged from looking at changes in the neural network's uncertainty in solving the inverse problem. 
Our approach is akin to, but distinct from, confusion-based techniques \cite{Nieuwenburg2017, lee2019confusion} which attempt to learn a model that directly predicts phase transitions but side-steps the requirement for ground truth labels by training with pseudo-labels.
}

\par{
Figure~\ref{fig:workflow} illustrates the steps of our approach.
In the first step (Fig.~\ref{fig:workflow}a top left panel), we performed high-fidelity phase-field simulations of pattern-forming processes in binary microstructures.
The intent of this first step is to generate a large and diverse set of microstructural patterns as a function of the process parameters (\textit{e.g.}, phase fraction, phase mobility, deposition rate) and therefore generate a database of transition regimes where the microstructural patterns switch from one class of patterns to another when the process parameters change.
As prototypical examples, we chose two \revision{pattern-forming} processes, namely the spinodal decomposition of a two-phase mixture and the formation of various self-organized microstructural patterns of binary alloys during the physical vapor deposition of thin films.
The first problem is chosen due to its simplicity in identifying different pattern-formation regimes as a function of the concentration and mobility of the two phases.
The second problem is used to test the generalizability of our approach, as it displays hard-to-\revision{discern} transitions \revision{from one class of patterns to another when the deposition process parameters change}.
\revision{The difficulty in this second problem lies in the identification of clear transitions in microstructural patterns which appear to have seemingly continuous and non-trivial changes of classes of patterns when the input process parameters vary.}
In the second step \revision{of our approach}, we represent these microstructural patterns in a latent space using a pre-trained convolutional neural network (CNN), namely ResNet-50 v2\cite{He2016}.
The goal of this step is to obtain a low-dimensional representation of these patterns (Fig.~\ref{fig:workflow}a right panel) \revision{in order to learn the structure of the data}.
While the ResNet model was trained on image data \revision{not} similar to our simulations, it is still able to discern complex shapes and motifs, which allows different microstructural patterns to be distinguishable.
Such performance shows the robustness of this step.
%
%We can use the same pre-trained \revision{ResNet} model to make the projection to the latent space for different types of data, as long as they can be displayed in a visual form.
%
%Additionally, the results are not specific to the choice of a particular type of model.
%
As \revision{described in the Supplementary Information (Supplementary Notes 1 and 2)}, we can obtain similar results using different pre-trained models.
In this work, we settled for ResNet-50 v2 \revision{as a prototypical CNN model} due to its wide accessibility and popularity.
The latent dimensions obtained from the ResNet model are subsequently regressed with a dense, feedforward, deep neural network to predict the original input process parameters that led to specific microstructural patterns.
This last step (Fig.~\ref{fig:workflow}a bottom left panel) can be seen as solving our inverse problem (\textit{i.e.} predicting the initial input process parameters from the output microstructural pattern).
The error between the prediction and ground truth process parameter is then compared on a validation set. 
Transitions between specific classes of microstructural patterns are identified by evaluating the sensitivity of the model prediction error (Fig.~\ref{fig:workflow}b) as a function of the input process parameters.
}
%%%%%%%%%%%%%%%%%%%%%%%%%%%%%%%%%%%%%%%%%%%%%%%%%%%%%%%%%%%%%%%%%
%%%%%%%%%%%%%%%%%%%%%%%%%%%%%%%%%%%%%%%%%%%%%%%%%%%%%%%%%%%%%%%%%
%%
%% RESULTS
%%
%%%%%%%%%%%%%%%%%%%%%%%%%%%%%%%%%%%%%%%%%%%%%%%%%%%%%%%%%%%%%%%%%
%%%%%%%%%%%%%%%%%%%%%%%%%%%%%%%%%%%%%%%%%%%%%%%%%%%%%%%%%%%%%%%%%
\section*{RESULTS}
%%%%%%%%%%%%%%%%%%%%%%%%%%%%%%%%%%%%%%%%%%%%%%%%%%%%%%%%%
%%% FIG 2: LATENT DESCRIPTION
%%%%%%%%%%%%%%%%%%%%%%%%%%%%%%%%%%%%%%%%%%%%%%%%%%%%%%%%%
 \begin{figure*}
        \centering
        \includegraphics[width=0.47\linewidth]{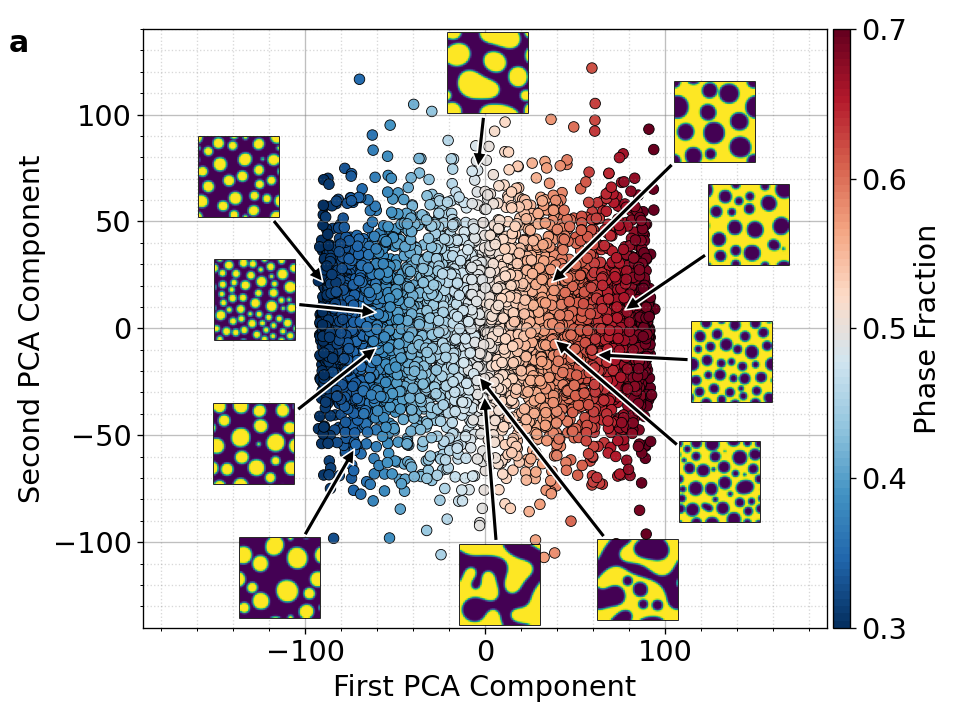}
        \includegraphics[width=0.47\linewidth]{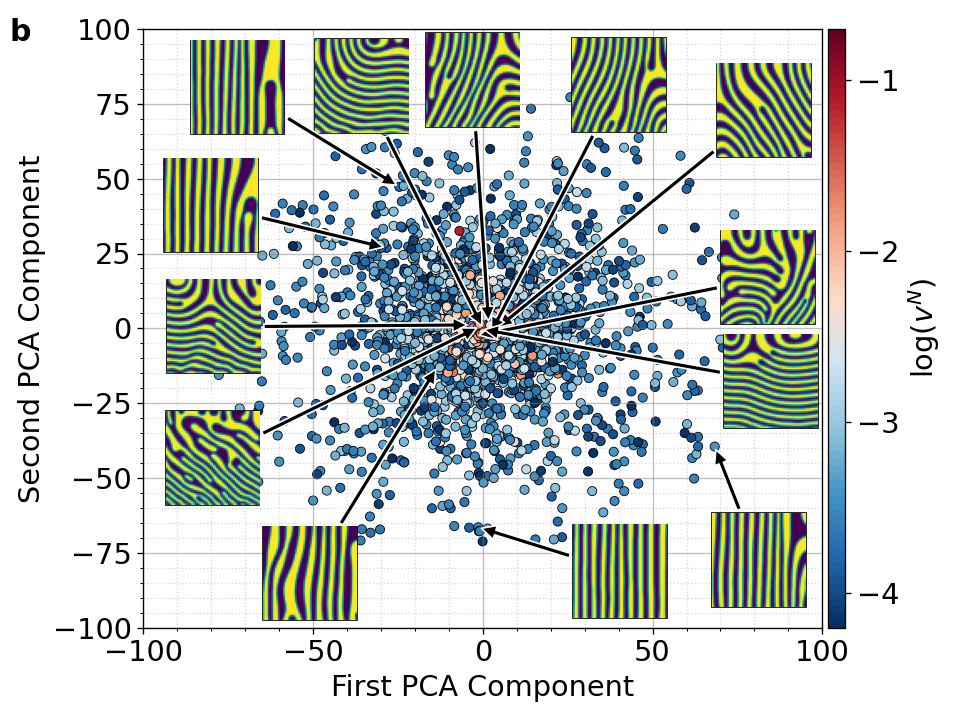}
        \includegraphics[width=0.47\linewidth]{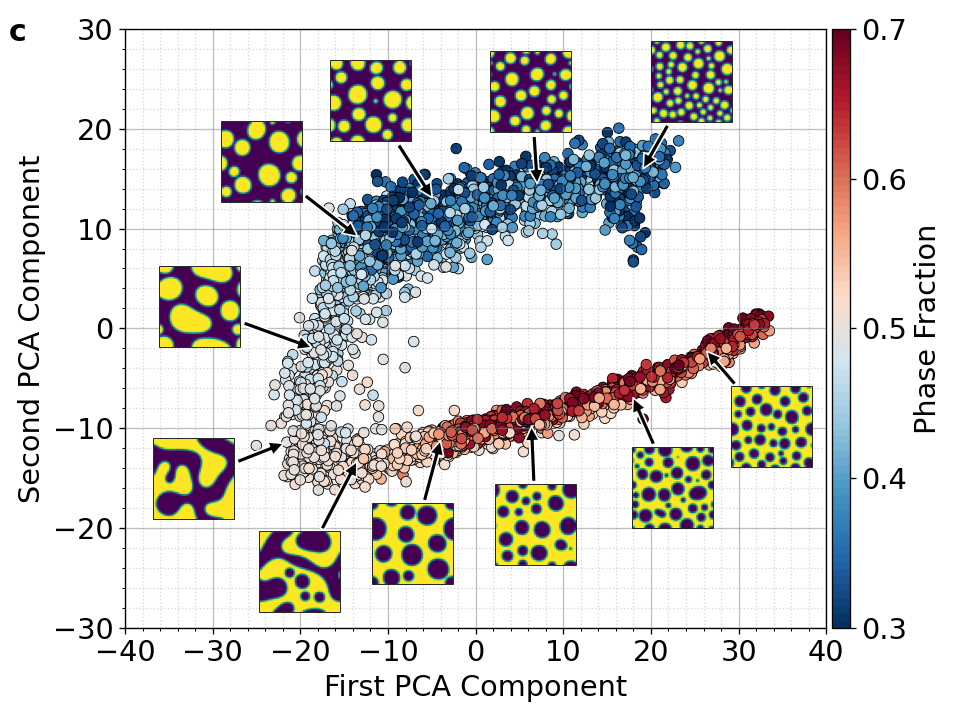}
        \includegraphics[width=0.47\linewidth]{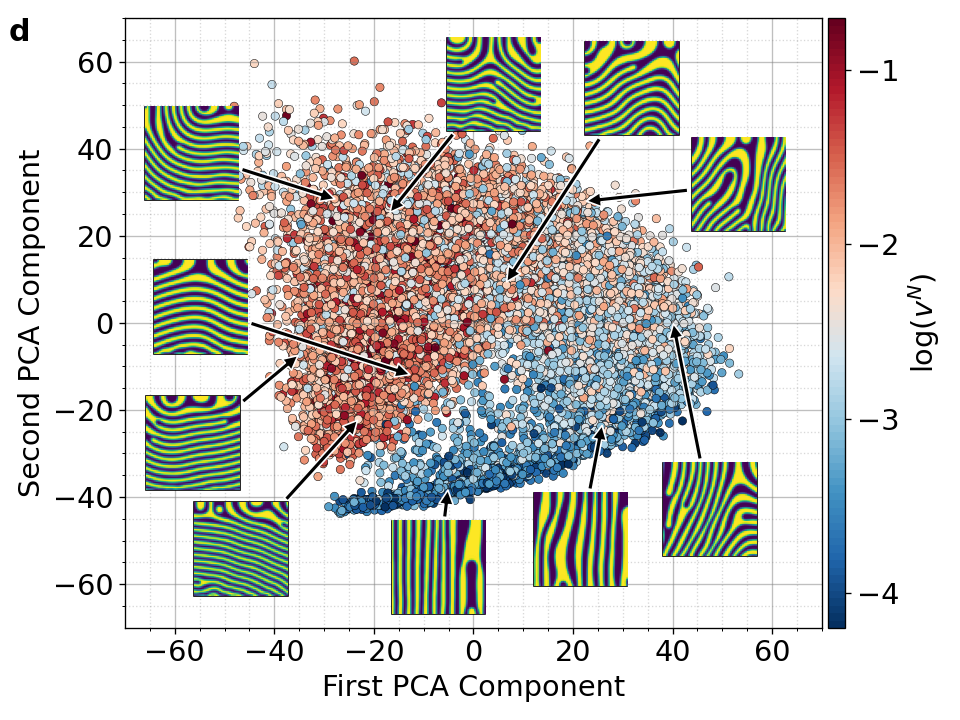}
        \includegraphics[width=0.47\linewidth]{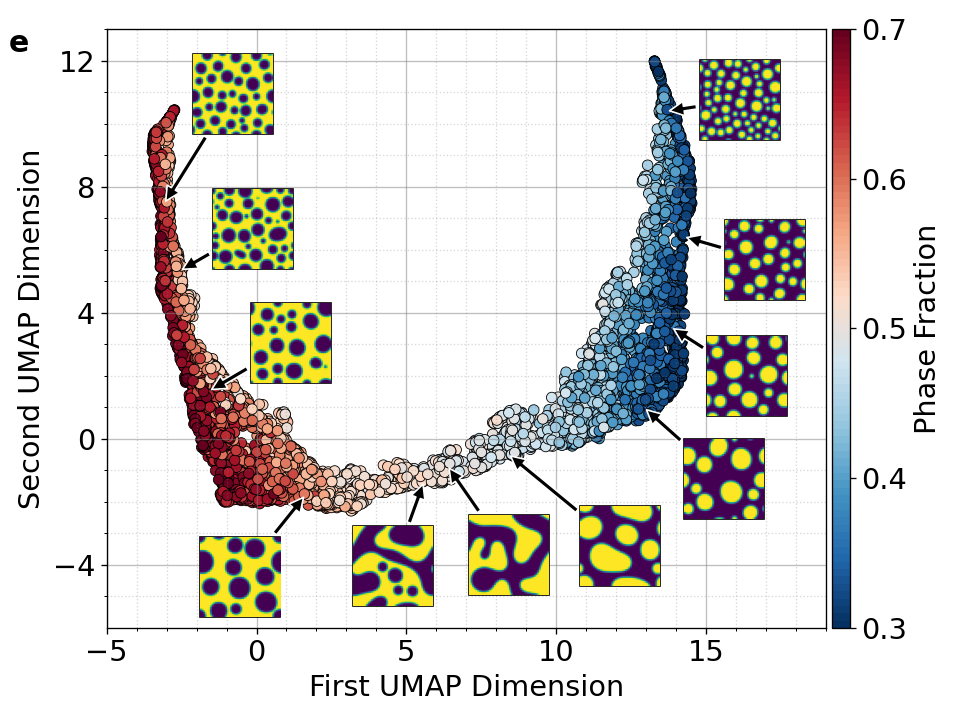}
        \includegraphics[width=0.47\linewidth]{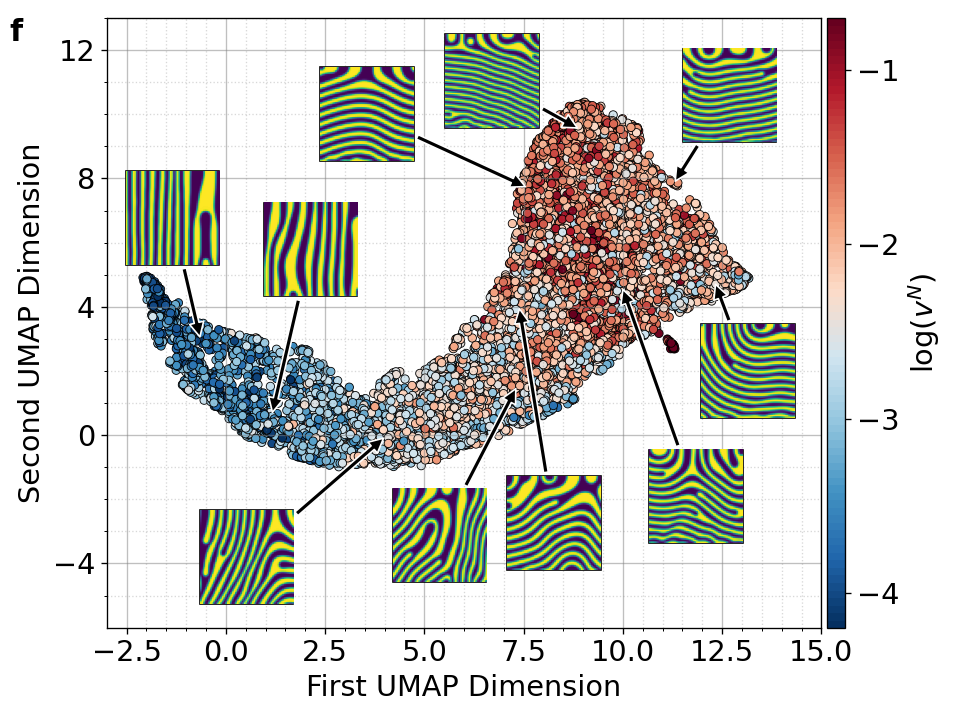}
        
        \caption{
        {\bf Latent representations of various microstructural patterns}. Spinodal decomposition of a two-phase mixture representations are displayed in left panels, 
        physical vapor deposition of a binary alloy thin film representations in right panels.
        {\bf a}-{\bf b} These panels show the projection using a linear embedding technique (PCA) directly on the microstructure images (defined as dimensionality reduction on the ambient space).
        {\bf c}--{\bf f} Second and third rows show the projection of the ResNet-50 v2 latent vector using PCA (panels {\bf c} and {\bf d}) and UMAP (panels {\bf e} and {\bf f}).
        Latent vectors are colored according to relevant input process parameters for both problems, namely the phase fraction ($f$) for spinodal decomposition and the normalized deposition rate ($v^{\rm{N}}$) for physical vapor deposition.
        Definition of the normalized deposition rate is provided in Methods.
        For the microstructure insets across both spinodal and physical vapor deposition problems, yellow denotes the A phase, and purple the B phase.
        }
        \label{fig:latent-representation}
 \end{figure*}
%%%%%%%%%%%%%%%%%%%%%%%%%%%%%%%%%%%%%%%%%%%%%%%%%%%%%%%%%
%%%
%%%%%%%%%%%%%%%%%%%%%%%%%%%%%%%%%%%%%%%%%%%%%%%%%%%%%%%%%
%%% FIG 3: PCA
%%%%%%%%%%%%%%%%%%%%%%%%%%%%%%%%%%%%%%%%%%%%%%%%%%%%%%%%%
 \begin{figure*}
        \centering
        \includegraphics[width=0.49\linewidth]{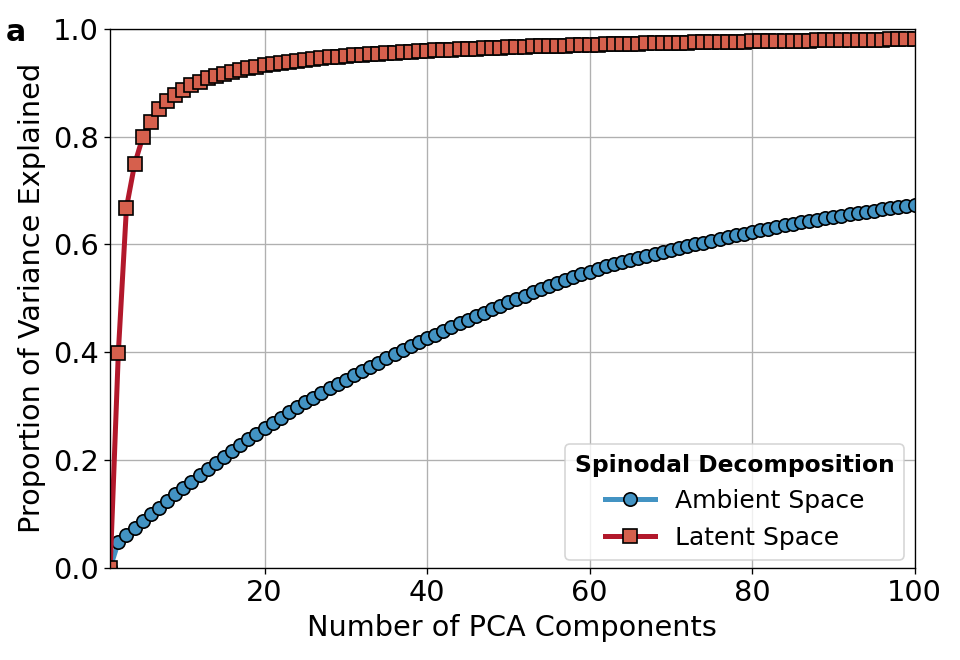}
        \includegraphics[width=0.49\linewidth]{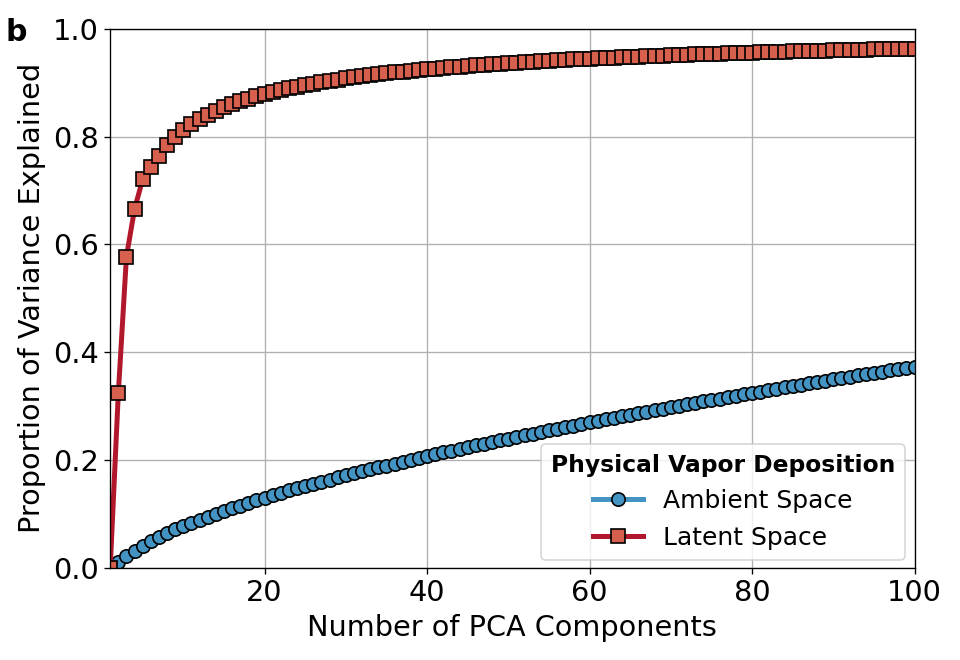}

        \caption{
        {\bf Cumulative explained variance versus ranked principal components when using PCA \revision{directly on actual microstructure images (ambient space) or on latent representation (latent space)}.}
        {\bf a} Explained variance for the spinodal decomposition problem.
        {\bf b} Explained variance for the physical vapor deposition problem.
        Ambient space denotes dimensionality reduction when PCA is performed directly on the high-dimensional microstructure images, while latent space denotes the projection when PCA is performed on the already compact representation of the microstructures obtained from the ResNet model.}
        \label{fig:explained_variance}
 \end{figure*}
%%%%%%%%%%%%%%%%%%%%%%%%%%%%%%%%%%%%%%%%%%%%%%%%%%%%%%%%%
%%%%%%%%%%%%%%%%%%%%%%%%%%%%%%%%%%%%%%%%%%%%%%%%%%%%%%%%%
%%%%%%%%%%%%%%%%%%%%%%%%%%%%%%%%%%%%%%%%%%%%%%%%%%%%%%%%%%%%%%%%%
%%%%%%%%%%%%%%%%%%%%%%%%%%%%%%%%%%%%%%%%%%%%%%%%%%%%%%%%%%%%%%%%%
%%
%% AMBIGUITY 
%%
%%%%%%%%%%%%%%%%%%%%%%%%%%%%%%%%%%%%%%%%%%%%%%%%%%%%%%%%%%%%%%%%%
%%%%%%%%%%%%%%%%%%%%%%%%%%%%%%%%%%%%%%%%%%%%%%%%%%%%%%%%%%%%%%%%%
 \subsection*{Ambiguity of identifying microstructural transitions}
 %In this section we discuss real space and latent embedding clustering. And difficulty of identifying transitions.
\par{
We first explore the ambiguity of fingerprinting nontrivial transitions in pattern formation when those patterns are represented in latent space.
We compare different dimensionality reduction techniques to represent the broad range of pattern configurations.
Details on the phase-field models and latent embedding approach are provided in Methods.
}

\par{
Figure~\ref{fig:latent-representation} illustrates how difficult and equivocal it is to identify transitions between different patterns as a function of the \revision{input} process parameters.
When taking the naive approach of performing principal component analysis~\cite{abdi2010principal} (PCA) directly on the original microstructure images for both problems (see panels a and b), we note that this projection technique is incapable of identifying or explaining transition regimes via distinct clusters.
This assertion is especially true when trying to distinguish the different microstructural patterns forming in the case of the physical vapor deposition model when the principal components are colored as a function of the normalized deposition rate $v^{\rm{N}}$ (defined as the deposition rate normalized by the average bulk phase mobility of the system, see exact definition in Methods).
}

\par{
\revision{To further illustrate this point}, we present the results for two \revision{additional} alternative projection techniques \revision{for both exemplar problems} in Fig.~\ref{fig:latent-representation} panels c--f.
In this case, even though the pre-trained ResNet-50 v2~\cite{He2016} model accurately learns a compact representation of the microstructural patterns (\textit{i.e.}, from a ${\rm 256}\times{\rm 256}$ pixelated microstructure representation to a 2048 latent vector), the dimension of the latent space is still too large to obtain any visual and interpretable classification of the different types of patterns forming in the two processes investigated in this study.
We used linear (PCA) and non-linear (uniform manifold approximation and projection~\cite{mcinnes2018umap, McInnes_2018b} or UMAP) embedding unsupervised algorithms to further \revision{reduce} the ResNet latent vectors into lower dimensions.
Figure~\ref{fig:explained_variance} shows the difference in explained variance when PCA is performed (i) directly on the actual high-dimensional microstructure images (denoted as ambient space) or (ii) on the already compact representation of the microstructures obtained from the ResNet model (denoted as latent space).
We note that combining PCA and ResNet requires fewer principal components to capture most of the variance in our datasets, making the dimensionality reduction procedure (PCA/UMAP+ResNet) appealing and useful.
}

\par{
Unfortunately, even with this improvement in the latent representation of the microstructural patterns, regardless of the embedding technique used in combination with ResNet, we observe that there is no intuitive, distinct clustering of data points about \revision{where exactly} the transitions \revision{are}.
This statement is apparent even in the simple case of the spinodal decomposition problem (Fig.~\ref{fig:latent-representation} panels c and e), where the transition is expected to occur for a 50/50 phase fraction.
When colored as a function of the phase fraction, $f$, the latent representation of the microstructural patterns continuously changes from one structure type to another (\textit{i.e}, from `B'-rich spherical-precipitate patterns when $f=0.3$ to `A'-rich spherical-precipitate patterns when $f=0.7$).
This latent representation does not clearly disentangle the different classes of patterns and cannot provide \textit{a priori} any information on where the transition occurs when the phase fraction changes.
The identification of topological transitions is doubly more difficult in the case of the physical vapor deposition problem (Fig.~\ref{fig:latent-representation} panels d and f).
In this case, neither PCA nor UMAP embedding show any obvious way to distinguish transitions between vertical-oriented to horizontal-oriented patterns, aside from noticing that nearby points will have similar microstructural patterns (as PCA and UMAP methods are designed to preserve similarities in feature space).
\revision{Here, in both examples, we used PCA/UMAP and the ResNet model for illustrative purposes only.
The results in Fig.~\ref{fig:latent-representation} demonstrate that identifying dynamic and second-order transitions using classical dimensionality reduction techniques does not enable us to unequivocally separate different pattern-formation regimes in latent space.
To further illustrate this point, as described in the Supplementary Information (Supplementary Note 1), we tested other projection methods (simple convolutional neural network, ResNet-34, t-SNE) to show that the results are not specific to the particular choice of a pre-trained model or dimensionality reduction technique.
These supplementary results show that we can obtain similar results using different pre-trained models.
}
}
%%%%%%%%%%%%%%%%%%%%%%%%%%%%%%%%%%%%%%%%%%%%%%%%%%%%%%%%%
%%% TABLE 1: EXAMPLES SPINODAL
%%%%%%%%%%%%%%%%%%%%%%%%%%%%%%%%%%%%%%%%%%%%%%%%%%%%%%%%%
    \begin{table*}[t]
    \centering
    \setlength{\tabcolsep}{4pt}
    \caption{\textbf{Prediction of mobility parameters in the case of the spinodal decomposition problem.}
    An example of instances with high and low sensitivity scores. For the microstructure insets, yellow denotes the A phase, and purple the B phase. The error $\Delta_i$ is defined as $\Delta_i=|\hat y - y|$, where $\hat y$ is the prediction and $y$ the true value. The sensitivity score is defined as $\mathcal{S}_i=\Delta_i^{-1}$. Instances 1042, 2106, 4157 belong to regime A as identified in Fig.~\ref{fig:spinodal-inversemapping}b. Instances 2868, 4496, 4545 belong to regime B as identified in Fig.~\ref{fig:spinodal-inversemapping}b. Instances 814, 1543, 4751 belong to regime C as identified in Fig.~\ref{fig:spinodal-inversemapping}b. Note that the high sensitivity scores (in bold) correspond to different input process parameters depending on the regime of importance. Each regime is separated by horizontal lines. 
    }
    \includegraphics[width=\textwidth]{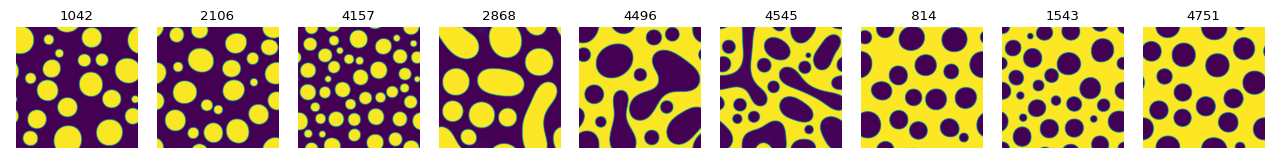}
    \begin{tabular}{cc|cccc|cccc}
        \\\hline \hline
        \multicolumn{2}{c|}{} & \multicolumn{4}{c|}{\bf {Phase A}} & \multicolumn{4}{c}{\bf {Phase B}} \\\hline
        Instance & Phase  & Predicted & Target  & Prediction & Sensitivity  & Predicted  & Target  & Prediction & Sensitivity \\
         Index &  Fraction  &  Mobility &  Mobility  &  Error ($\Delta_i$) & Score ($\mathcal{S}_i$)  & Mobility & Mobility  & Error ($\Delta_i$) & Score ($\mathcal{S}_i$)\\\hline\hline
        1042 & 0.34766  & 0.3758 & 0.9408 & 0.5650 & \phantom{0}1.7700 & 0.4418 & 0.5062 & 0.0644 & {\bf 15.5344}\\%\hline  
        2106 & 0.34964 & 0.3671 & 0.1338 & 0.2332 & \phantom{0}4.2874 & 0.2488 & 0.3725 & 0.1237 & {\bf \phantom{0}8.0867}\\%\hline
        4157 & 0.35344 & 0.4290 & 0.3608 & 0.0682 & 14.6568 & 0.1188 & 0.0979 & 0.0209 & {\bf 47.8391}\\\hline
        2868 & 0.49802 & 0.3062 & 0.8484 & 0.5422 & \phantom{0}1.8443 & 0.3434 & 0.2021 & 0.1413 & {\bf \phantom{0}7.0785}\\%\hline
        4496 & 0.50287 & 0.2445 & 0.0432 & 0.2013 & {\bf \phantom{0}4.9672} & 0.3609 & 0.7487 & 0.3878 & 2.5784\\%\hline
        4545 & 0.50289 & 0.1002 & 0.0464 & 0.0538 & 18.6006 & 0.2371 & 0.2705 & 0.0333 & {\bf 29.9854}\\\hline
        \phantom{0}814 & 0.64599 & 0.4386 & 0.3852 & 0.0533 & {\bf 18.7477} & 0.4604 & 0.3713 & 0.0891 & 11.2179\\%\hline
        1543 & 0.64969 & 0.2343 & 0.2529 & 0.0186 & {\bf 53.6827} & 0.4152 & 0.3636 & 0.0516 & 19.3968\\%\hline
        4751 & 0.65144 & 0.4143 & 0.5127 & 0.0984 & {\bf 10.1611} & 0.4619 & 0.8399 & 0.3780 & \phantom{0}2.6454\\\hline\hline
    \end{tabular}
    \label{tab:spinodal_decomposition_examples}
    \end{table*}

%%%%%%%%%%%%%%%%%%%%%%%%%%%%%%%%%%%%%%%%%%%%%%%%%%%%%%%%%%%%%%%%%
%%%%%%%%%%%%%%%%%%%%%%%%%%%%%%%%%%%%%%%%%%%%%%%%%%%%%%%%%%%%%%%%%
%%
%% SOLVING INVERSE MAPPING
%%
%%%%%%%%%%%%%%%%%%%%%%%%%%%%%%%%%%%%%%%%%%%%%%%%%%%%%%%%%%%%%%%%%
%%%%%%%%%%%%%%%%%%%%%%%%%%%%%%%%%%%%%%%%%%%%%%%%%%%%%%%%%%%%%%%%%
 \subsection*{Solving the inverse mapping problem to identify transition regimes}
\par{
Because the seemingly obvious break down of automatically clustering and classifying microstructural patterns for the examples shown in Fig\@.~\ref{fig:latent-representation}, and inspired by the universality principle for dynamical systems (\textit{i.e.}, near a phase transition the properties of a system become nearly independent of some dynamical details of that system), we instead develop a method that aims at measuring the difficulty of solving an inverse problem: mapping the microstructural patterns back to the original input process parameters.
If this task is accurate, then the input process parameters uniquely map to a given microstructural pattern.
Conversely, if this mapping is difficult or ambiguous, then there is a broad range of \revision{possible} microstructural patterns \revision{that can be obtained} with similar input process parameters, possibly indicating a topological transition (similar to the changes in sensitivity to some parameters when reaching the critical universality classes in second-order phase transitions~\cite{stauffer1972universality, nikoghosyan2016universality}).
This inverse-mapping solution is outlined in Fig\@.~\ref{fig:workflow}. Details on the inverse mapping workflow are provided in Methods.
\revision{Details on the quality of the training of the neural network are also provided in Methods and in the Supplementary Information, Supplementary Note 5.}
}
   
\par
\begin{figure*}
        \includegraphics[width=\linewidth]{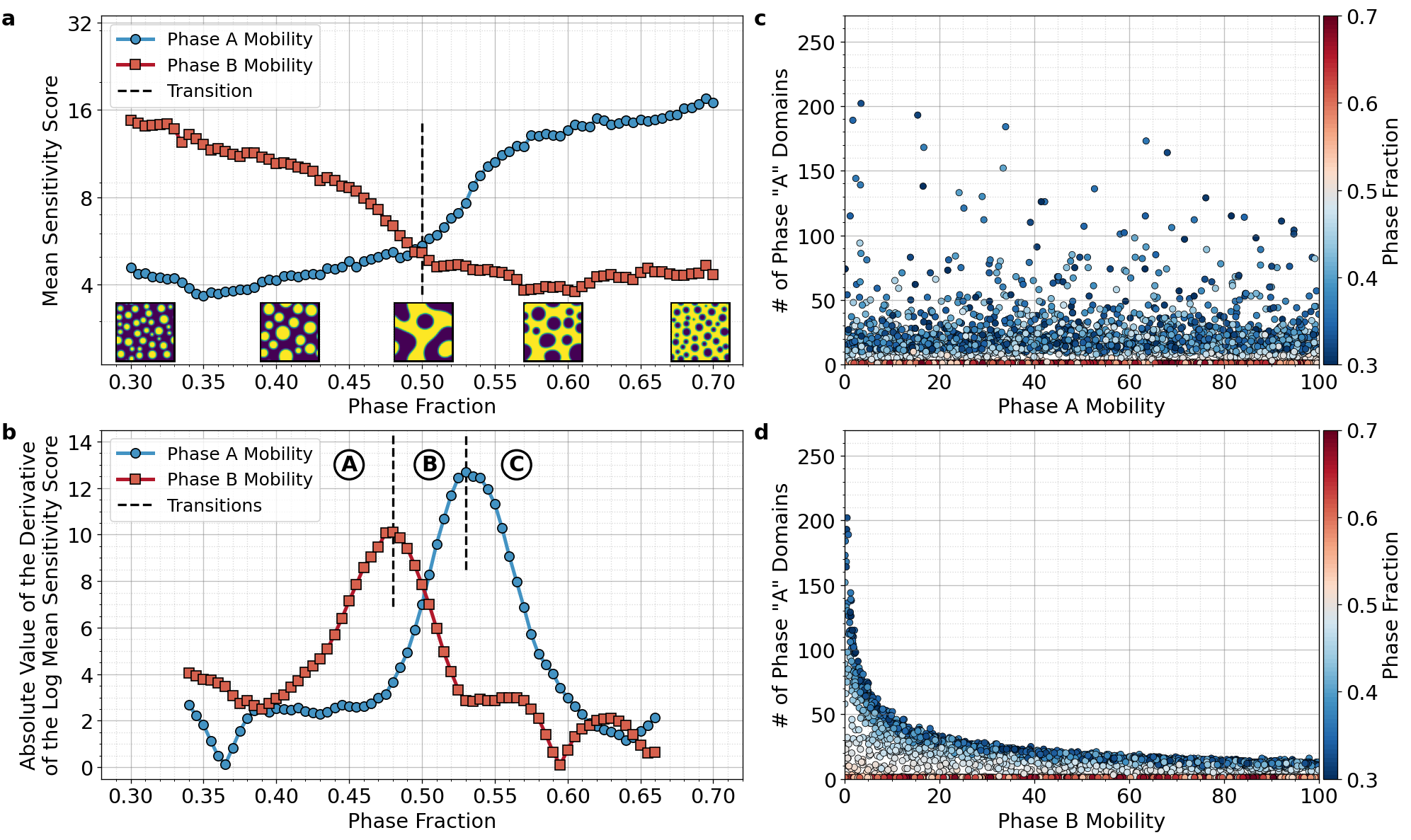}

        \caption{{\bf Predicting the process parameters for spinodal deposition of a two-phase mixture.}
        {\bf a} The mean sensitivity score for predicting mobilities as a function of phase fraction. When changing the fraction from low to high, the dominant role of the phase mobility switches from B to A. The dashed line indicated the expected transition from a `A'-rich patterns regime to a `B'-rich patterns regime.
        Insets depict example simulation realizations for particular values of the phase fraction order parameter. For the microstructure insets, yellow denotes phase A, and purple phase B.
        {\bf b} Variation (derivative) of the sensitivity score as a function of phase fraction.  
        {\bf c}--{\bf d} Scatterplots of the variability on the number of phase domains A as a function of the phase A and B mobilities respectively.
        }
        \label{fig:spinodal-inversemapping}
 \end{figure*}

{
We first illustrate how intuitively this workflow works for the spinodal decomposition problem, because in this simple case, based on symmetry considerations of the free energy functional, we expect the transition to occur at the 50\% phase fraction.
Examples of predictions of input process parameters with high and low errors are listed in Table \ref{tab:spinodal_decomposition_examples}.
We present results for the full range of parameters in Fig\@.~\ref{fig:spinodal-inversemapping}.
To obtain those results, we calculated the average mean absolute error (MAE) of our predictions in intervals $(f-0.025, f+0.025)$ where $f$ denotes the phase fraction.
The sensitivity score per instance ``$i$'' is defined as the inverse of the MAE for each value of phase fraction, $\mathcal{S}_{i} = 1/|\hat y_i - y_i|$, where $\hat y$ is the prediction and $y$ the true value.
The mean sensitivity score of the entire dataset in a given interval is defined as $\mathcal{S} = N/\sum_i^N|\hat y_i - y_i|$.
A high sensitivity score indicates that we are able to predict the input process parameters accurately and \revision{this parameter acts} as an indicator for a given class microstructural pattern.
\revision{Conversely,} a low sensitivity score indicates that the relation between the initial value of the process parameter and the microstructural pattern is weak, \revision{and this process parameter has little influence on the formation of that specific pattern}.
When the sensitivity score changes from high to low or vice versa, \revision{this indicates that we are near a transition with respect to that input process parameter}.
In other words, the task of evaluating the change in sensitivity score is analogous to a mutual information analysis~\cite{gierlichs2008mutual} from the field of information theory, which consists of evaluating the information gain to feature selection.
\revision{In this context, here the underlying assumption is that, for a given set of process parameters, the microstructural pattern formation is a stochastic process, and each microstructure realization is a realization of this stochastic process.
Mutual information here is calculated between the microstructural patterns and the process parameters and measures the change in uncertainty for the process parameters given a known observed microstructure.}
}

{
We observe from the results listed in Table \ref{tab:spinodal_decomposition_examples} that for high phase fractions (\textit{i.e.}, when A is the majority phase, $f>0.5$) the outcome is highly sensitive to the choice of initial value of the phase A mobility.
Conversely, the outcome does not change in any meaningful way if we vary the value of phase B mobility (as indicated by low values of the sensitivity score calculated with respect to the phase B mobility).
An interesting observation is that looking at the highest sensitivity scores, we can distinguish three groups of patterns.
Instances 1042, 2106, and 4157 display high sensitivity scores for the phase B mobility.
Similarly, instances 814, 1543, and 4753 display high sensitivity scores for the phase A mobility.
However, instances 2868, 4496, and 4545 show mixed results with sometimes high sensitivity score for phase A or phase B mobilities.
While individual sensitivity scores may already be good predictors of when the topological \revision{transition} may occur, we observe a large variability in those scores.
For example, comparing individual sensitivity scores for similar microstructural patterns in Table~\ref{tab:spinodal_decomposition_examples}, the sensitivity score varies from 1 (instance 1042) to 14 (instance 4157).
\revision{Therefore, to detect the transitions we will look at the averaged value of the sensitivity score, as discussed next.}
}
%%%%%%%%%%%%%%%%%%%%%%%%%%%%%%%%%%%%%%%%%%%%%%%%%%%%%%%%%
%%% FIG 4: SPINODAL INVERSE MAPPING
%%%%%%%%%%%%%%%%%%%%%%%%%%%%%%%%%%%%%%%%%%%%%%%%%%%%%%%%%
%%%%%%%%%%%%%%%%%%%%%%%%%%%%%%%%%%%%%%%%%%%%%%%%%%%%%%%%%
%%%%%%%%%%%%%%%%%%%%%%%%%%%%%%%%%%%%%%%%%%%%%%%%%%%%%%%%%

\par{
%individual (per-instance) scores
Comparing individual scores to identify microstructural transitions could be deceiving ($\mathcal{S}_{i}$ with respect to phase A mobility for example 4157 is larger than $\mathcal{S}_{i}$ with respect to phase B mobility for example 2106).
We can mitigate this issue by evaluating the mean sensitivity scores on the predicted mobilities.
The mean sensitivity scores $\mathcal{S}$ on the predicted mobilities for phases A and B are presented in Fig\@.~\ref{fig:spinodal-inversemapping}a.
Variations (\textit{i.e.} derivative) of the sensitivity scores with respect to the \revision{phase fraction} are presented in Fig\@.~\ref{fig:spinodal-inversemapping}b.
To see whether or not we can identify the transition regime at the 50\% phase fraction, we can first look at the left part of Fig\@.~\ref{fig:spinodal-inversemapping}a.
First, we would expect a strong correlation between \revision{the value of the} phase B mobility and the size or shape of the phase A domains \revision{when} B is the majority phase.
\revision{This correlation translates into a high value of} the sensitivity score with respect to phase B mobility.
Conversely, we would also expect a weak relation between \revision{the value of the} phase B mobility and the dynamics of the A domain when A is the majority phase, \textit{i.e.} in this case the sensitivity score will be low.
Indeed, if we look at the data and plot the number of phase A domains (\textit{i.e.} the number of \revision{small spherical domains}) as a function of mobility (see Fig.~\ref{fig:spinodal-inversemapping} panels c--d), we see that the relation to phase B mobility is very strong (especially for low values of phase fraction), while relation to phase A mobility is almost non-existent (it does not matter what the value of phase A mobility is, the number of domains might be very different).
As such, we note that when predicting the mobilities for phases A and B respectively in Fig\@.~\ref{fig:spinodal-inversemapping}a, the sensitivity of those predictions switches to a relatively low sensitivity score at the 50/50 phase fraction.
This is further evidenced if we look at the variation of the sensitivity score as a function of the \revision{phase fraction} in Fig.~\ref{fig:spinodal-inversemapping}b.
We observe a transition from (i) a regime with a monomodal microstructure of A-rich \revision{spherical} precipitates entrapped in a B matrix (denoted as encircled A in Fig\@.~\ref{fig:spinodal-inversemapping}b) to (ii) an intermediate microstructural configuration (denoted as encircled B) which consists of interconnected A or B domains to (iii) finally another regime (denoted as encircled C) with a monomodal microstructure of B-rich \revision{spherical} precipitates entrapped in a A matrix. 
This intermediate regime corresponds to the gradual transition in pattern formation between the two other distinct regimes.
}
%%%%%%%%%%%%%%%%%%%%%%%%%%%%%%%%%%%%%%%%%%%%%%%%%%%%%%%%%%%%%%%%%
%%%%%%%%%%%%%%%%%%%%%%%%%%%%%%%%%%%%%%%%%%%%%%%%%%%%%%%%%%%%%%%%%
%%
%% GENERALIZATION TO COMPLEX PROCESSES 
%%
%%%%%%%%%%%%%%%%%%%%%%%%%%%%%%%%%%%%%%%%%%%%%%%%%%%%%%%%%%%%%%%%%
%%%%%%%%%%%%%%%%%%%%%%%%%%%%%%%%%%%%%%%%%%%%%%%%%%%%%%%%%%%%%%%%%    
\subsection*{Generalization to more complex pattern-forming processes: physical vapor deposition of binary alloy thin films} 
\par{
We now turn to the physical vapor deposition problem which presents multiple transition regimes when the deposition rate and bulk mobility change.
At low normalized deposition rates (corresponding in the present case to values of $\log(v^{\rm N}) \sim -4$), vertical-oriented microstructural patterns form, while higher normalized deposition rates (corresponding in the present case to values of $\log(v^{\rm N}) \sim -1$) yield horizontal layer patterns.
At very high normalized deposition rates (corresponding in the present case to values of $\log(v^{\rm N}) \sim -0.5$) random microstructural patterns can form.
Similarly to the spinodal decomposition problem above, we list in Table~\ref{tab:pvd_examples} examples of predictions with high and low sensitivity scores for the deposition rate ($|{\bf{v}}|$).
Predictions of the deposition rate, $|\bf{v}|$, for vertical-oriented patterns are highly sensitive (\textit{i.e.} $\mathcal{S}_{i}$ is high) at low normalized deposition rates ($\log(v^{\rm N}) \sim -4.0$), while predictions for the average bulk mobility, $M^{\rm Bulk}$, are relatively insensitive.
We observe a different behavior for the normalized deposition rates that result in horizontal layer patterns for which in this case predictions for the deposition rate exhibit a low sensitivity score (see instance 444) and a high sensitivity score for the average bulk mobility.
}
%%%%%%%%%%%%%%%%%%%%%%%%%%%%%%%%%%%%%%%%%%%%%%%%%%%%%%%%%
%%% TABLE 2: EXAMPLE PVD
%%%%%%%%%%%%%%%%%%%%%%%%%%%%%%%%%%%%%%%%%%%%%%%%%%%%%%%%%
    \begin{table*}[t]
    \centering
    \setlength{\tabcolsep}{4pt}
    \caption{\textbf{Prediction of mobility parameters in the case of the physical vapor deposition problem.}
    An example of instances with high and low sensitivity score. Instances 1844, and 6316 belong to regime A as identified in Fig\@.~\ref{fig:pvd-inversemapping}c. Instances 15812 and 7304 belong to regime B as identified in Fig\@.~\ref{fig:pvd-inversemapping}c. Instances 1492 and 444 belong to regime C as identified in Fig\@.~\ref{fig:pvd-inversemapping}c. Instances 1752 and 3772 belong to regime D as identified in Fig\@.~\ref{fig:pvd-inversemapping}c.
    Each regime is separated by horizontal lines.
    }
    \includegraphics[width=\textwidth]{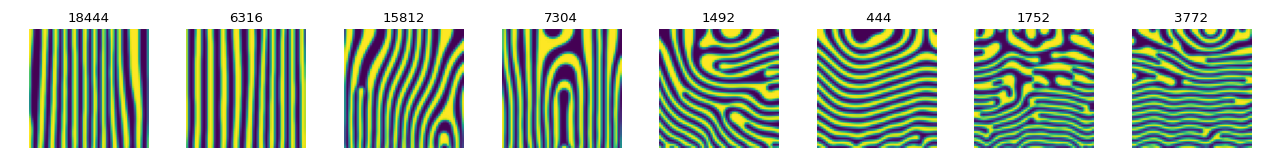}
    \begin{tabular}{cccc|ccc|ccc}
        \\\hline \hline
        \multicolumn{4}{c|}{} & \multicolumn{3}{c|}{\bf Deposition Rate} & \multicolumn{3}{c}{\bf Average Bulk Mobility} \\\hline
        Instance & Deposition  & Deposition & Avg. Surface  & Predicted & Target & Sensitivity  & Predicted & Target  & Sensitivity \\
        Index & $\log\left(v^N\right)$  & Angle & Mobility  & Deposition & Deposition & Score ($S_i$) & Mobility & Mobility  & Score ($S_i$)\\\hline\hline
        18444 & -4.049 & 59.776 & 38.060 &  0.204 &  0.118 & 11.665 &  4.853 &  5.852 &  1.001\\
        \phantom{0}6316 & -4.014 & 45.214 & 32.311 &  0.214 &  0.190 & 41.314 &  5.430 &  7.467 &  0.491\\\hline
        15812 & -3.006 & 61.088 & 15.890 &  0.316 &  0.264 & 19.311 &  4.087 &  4.680 &  1.686\\
        \phantom{0}7304 & -2.993 & 62.599 & \phantom{0}6.886 &  0.286 &  0.173 &  8.865 &  4.794 &  3.070 &  0.580\\\hline
        \phantom{0}1492 & -2.026 & 47.498 & 19.885 &  0.773 &  0.967 &  5.150 &  3.514 &  5.406 &  0.529\\
        \phantom{00}444 & -1.990 & 78.487 & 18.747 &  0.581 &  0.737 &  6.406 &  3.857 &  5.282 &  0.702\\\hline
        \phantom{0}1752 & -1.032 & 82.301 & 19.825 &  0.748 &  0.797 & 20.576 &  1.965 &  2.215 &  4.011\\
        \phantom{0}3772 & -0.993 & 79.439 & 12.418 &  0.838 &  0.980 &  7.018 &  2.460 &  2.601 &  7.063\\
        \hline\hline
    \end{tabular}
    \label{tab:pvd_examples}
    \end{table*}
%%%%%%%%%%%%%%%%%%%%%%%%%%%%%%%%%%%%%%%%%%%%%%%%%%%%%%%%%
%%%%%%%%%%%%%%%%%%%%%%%%%%%%%%%%%%%%%%%%%%%%%%%%%%%%%%%%%

\par
\begin{figure*}
        \centering
        \includegraphics[width=0.495\linewidth]{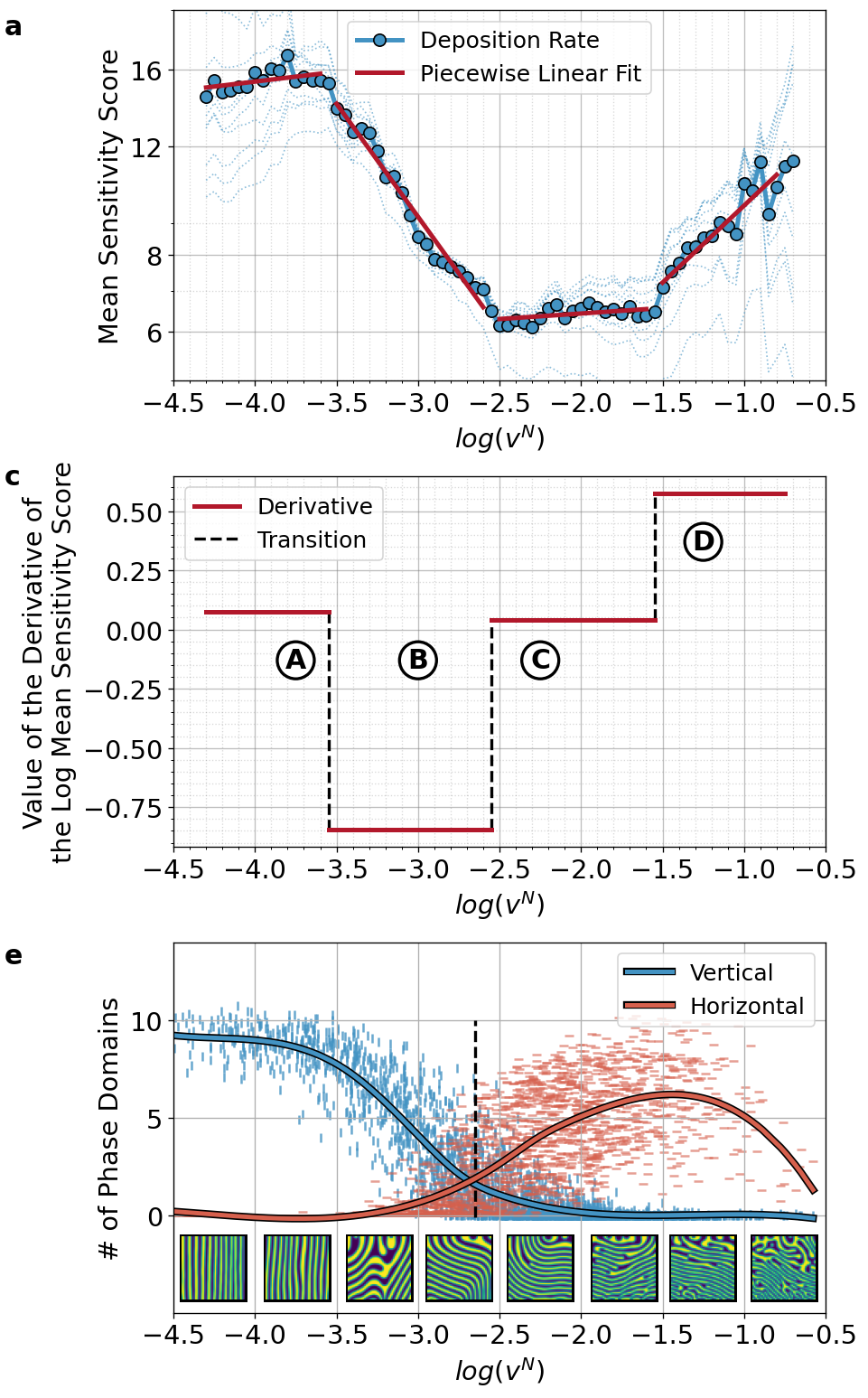}
        \includegraphics[width=0.495\linewidth]{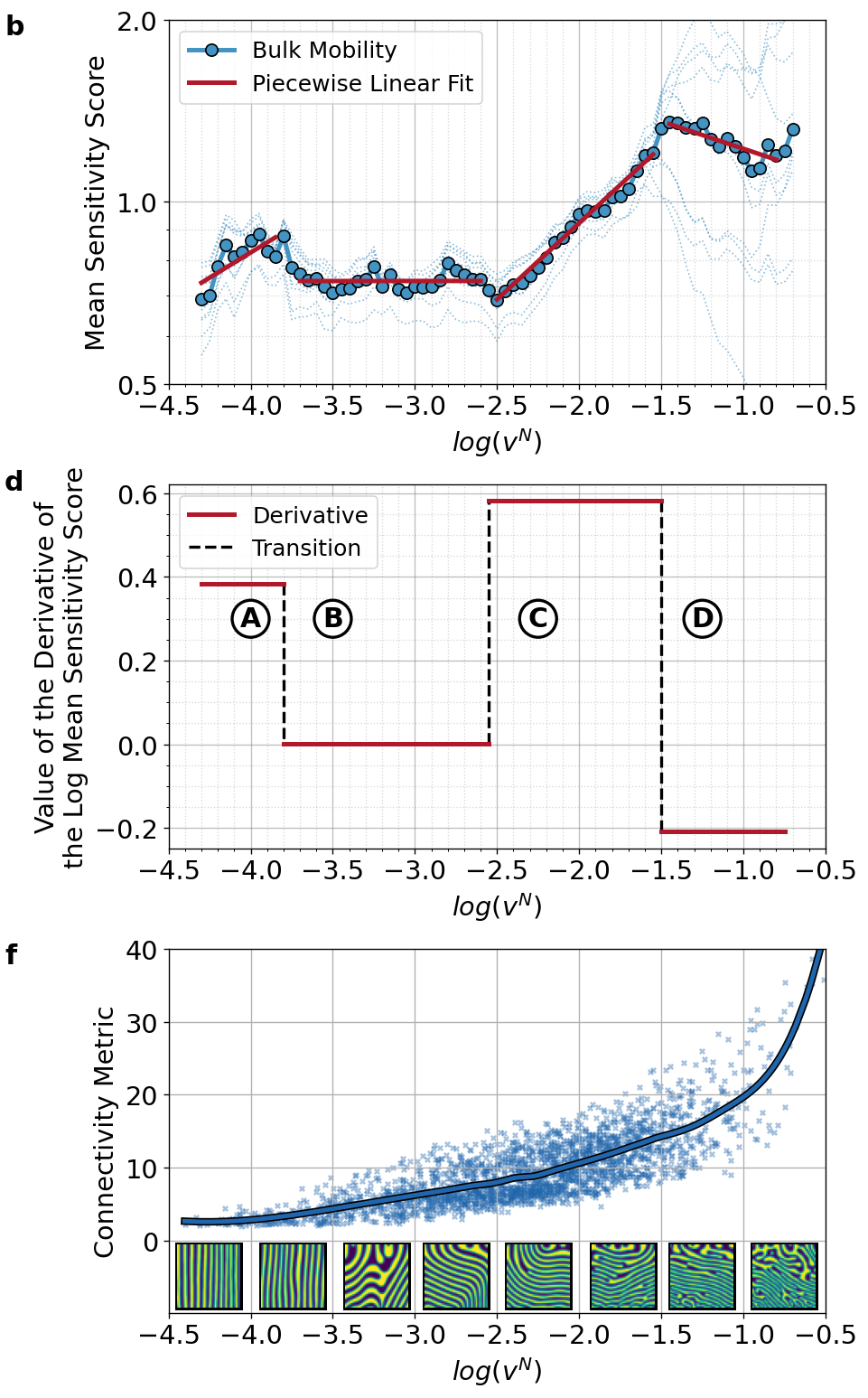}
        \caption{{\bf Identifying microstructural topological transitions for the physical vapor deposition of thin films.}
        {\bf a} Sensitivity score for predicting the deposition rate ($|\bf{v}|$) as a function of the normalized deposition rate.
        {\bf b} Sensitivity score for predicting the bulk mobility ($M^{\rm{Bulk}}$) as a function of the normalized deposition rate.
        {\bf c} Variation (derivative) of the sensitivity score for the deposition rate as a function of the normalized deposition rate.
        {\bf d} Variation (derivative) of the sensitivity score for predicting the bulk mobility as a function of the normalized deposition rate.
        {\bf e} Number of horizontal and vertical phase domains as a function of the normalized deposition rate.
        {\bf f} Topology connectivity (defined as the number of irregular elements of morphologies \textit{e.g.}, kinks, terminated stripes, \textit{etc.}) as a function of the normalized deposition rate.
        Thin blue lines in {\bf a} and {\bf b} illustrates particular runs of our procedure and indicate the variability around the averaged sensitivity score.}
        \label{fig:pvd-inversemapping}
 \end{figure*}

{
We show in Fig\@.~\ref{fig:pvd-inversemapping} how solving the inverse mapping problem for predicting the deposition rate, $|{\bf v}|$ (Fig\@.~\ref{fig:pvd-inversemapping} panels a, c), and average bulk mobility, $M^{\rm Bulk}$ (Fig\@.~\ref{fig:pvd-inversemapping} panels b, d), helps us identify the transition regime(s) between vertical- and horizontal-oriented microstructural patterns as a function of the normalized deposition rate $v^{\rm N}$.
First, we notice in Fig\@.~\ref{fig:pvd-inversemapping} panels a and b that when the orientation of the microstructural patterns changes from vertically oriented (denoted as regime A) to horizontally oriented (denoted as regime C), the sensitivity score for the deposition rate gradually changes from high to low accordingly and from low to high for the average bulk mobility.
We can interpret it as a change in the importance of the input process parameter based on the resulting microstructural pattern considered (horizontal \textit{vs}\@.~vertical).
The piecewise linear fits in Fig\@.~\ref{fig:pvd-inversemapping} panels a and b provide a visual guide as to when a transition occurs, while the thin blue lines show individual instances and the resulting variability around the averaged  sensitivity scores.
While the sensitivity scores on the deposition rate and average bulk mobility enable us to detect the major change in the overall orientation of the microstructural patterns, the variations in the level of the sensitivity scores (\textit{cf}\@.~Fig\@.~\ref{fig:pvd-inversemapping}c) reveal much more subtle changes.
Our self-supervised approach detects \revision{a second transition and the existence of} an intermediate regime (denoted by encircled B) that corresponds to a regime with a class of microstructures that are neither vertical nor horizontal.
We also note a \revision{third transition} at $\log v^{\rm N} \sim -1.5$ which corresponds to a horizontal-to-multimodal (or horizontal-to-random) transition.
When we compare the variation of the sensitivity scores between the two input process parameters in Fig\@.~\ref{fig:pvd-inversemapping} \revision{panels} c and d, we confirm the identification of those four regimes.
The marginal differences as to when the transitions occur in terms of the normalized deposition rate illustrate the relative importance of the specific predicted input process parameters on those regimes.
Second, our results reveal the ``range'' of each transition \revision{and existence of a specific pattern regime}.
Indeed, we observe for instance that the horizontal-oriented pattern regime only exists for a specific range of normalized deposition rates, namely for $\log v^{\rm N} \in [-2.5,-1.5]$.
}
%
%
%%%%%%%%%%%%%%%%%%%%%%%%%%%%%%%%%%%%%%%%%%%%%%%%%%%%%%%%%
%%% FIG 5: PVD INVERSE MAPPING
%%%%%%%%%%%%%%%%%%%%%%%%%%%%%%%%%%%%%%%%%%%%%%%%%%%%%%%%%
%%%%%%%%%%%%%%%%%%%%%%%%%%%%%%%%%%%%%%%%%%%%%%%%%%%%%%%%%
%%%%%%%%%%%%%%%%%%%%%%%%%%%%%%%%%%%%%%%%%%%%%%%%%%%%%%%%%

\par{
When plotting the number of horizontal- and vertical-oriented microstructural domains as a function of the normalized deposition rate in Fig\@.~\ref{fig:pvd-inversemapping}c, we confirm that indeed the microstructure topology switches from one type of pattern to the other for the detected transitions via our self-supervised approach.
When plotting these regimes as a function of the connectivity of the pattern in Fig\@.~\ref{fig:pvd-inversemapping}d, we note that the first transition (from A to B) can be associated with a transition from ordered vertical-oriented patterns to disjointed, more complex vertical patterns.
The second transition (from B to C) then corresponds to a clear change from disjointed vertical patterns to horizontal layer patterns.
The last transition (from C to D) corresponds to a transition from monomodal horizontal microstructure to a multimodal patterned structure.
These results contrast with those obtained via a human-annotated method to map two-phase patterned morphologies~\cite{Lu2012} for a similar problem, although it is unclear if this previous work fully captures the nuanced morphology classes.
}
%
%%%%%%%%%%%%%%%%%%%%%%%%%%%%%%%%%%%%%%%%%%%%%%%%%%%%%%%%%%%%%%%%%
%%%%%%%%%%%%%%%%%%%%%%%%%%%%%%%%%%%%%%%%%%%%%%%%%%%%%%%%%%%%%%%%%
%%
%% DISCUSSION
%%
%%%%%%%%%%%%%%%%%%%%%%%%%%%%%%%%%%%%%%%%%%%%%%%%%%%%%%%%%%%%%%%%%
%%%%%%%%%%%%%%%%%%%%%%%%%%%%%%%%%%%%%%%%%%%%%%%%%%%%%%%%%%%%%%%%%
\section*{DISCUSSION}
\par{
We have demonstrated \revision{above} that solving the inverse problem of mapping microstructural patterns back to the corresponding input process parameters is a powerful approach to infer and characterize transitions in pattern-forming processes, without any \textit{a priori} knowledge of where these \revision{transitions} occur in the input space.
This approach therefore allows microstructures to be objectively determined rather than relying on human perception and \revision{predefined} labelling.
In contrast to using clustering as a mean to identify transitions, this approach relies on the ability to relate transitions in microstructural patterns \revision{to} changes in the error of predicting the inverse problem of mapping microstructure to input process parameters.
This self-supervised approach relaxes the assumption in confusion-based techniques\cite{Nieuwenburg2017} which requires two distinctive phases to exist in the considered transition regime.
Instead, if there are multiple, intermediate, \revision{or hard-to-discern} transitions present in the data (including \revision{transitions} not characterized as first- or second-order transitions, \textit{cf}.\ Fig\@.~\ref{fig:pvd-inversemapping}), we can still detect and identify \revision{where these transitions occur and establish a hierarchy of transitions via our approach}.
%
%Some transition regimes can not be characterized as a simple first- or second-order transition between some distinct, well-defined classes.
%
As seen in the physical vapor deposition example, transitions \revision{can} happen at different levels of complexity (\textit{i.e.}\ monomodal \textit{vs}.\ multimodal patterns as seen in panels c and d of Fig.~\ref{fig:pvd-inversemapping}).
Therefore, we can use this approach to provide a hierarchy of transitions from the most pronounced (such as pattern orientation) to more subtle (such as complexity).
\revision{The CNN model used in this work can detect both easy geometric shapes (such as the orientation and width of the domains, size of the subdomains, \textit{etc}.) as well as more complex patterns (bifurcation of different subdomains).
This ability is possible due to the way information is organized in large CNN models, such as the ResNet model used in the present work.
Even if the model is initially trained on images from different scientific domains (animals, plants, and various natural objects such as cars and buildings), the neural network learns to recognize basic patterns such as the orientation of lines, curvatures of edges, \textit{etc}.
Those simple patterns are encoded in the lower part of the network and are easily transferable between domains.
This remarkable property of CNNs to generalize~\cite{bau2020understanding} from one visual domain to another allowed us to efficiently apply a model that was pre-trained on the ImageNet database to our pattern-forming process data sets.}
One possible improvement to provide even more granularity \revision{and generality} in detecting and categorizing \revision{transition} sub-regimes within these pattern-forming processes would be to look at cross-correlations between sensitivity scores or consider the sensitivity scores directly as a multi-dimensional surface response of the input process parameter space.
Local and global minima within this response could be identified as the different transitions.
}

\par{
\revision{In this study, we passed microstructure images to an existing, pre-trained ResNet model.
We did not have to fine-tune the ResNet model.
We kept this model as is to show how well the framework generalizes from one domain to another (in other words, how robust the framework is to the choice of the raw data embedding method).
To further support this claim, we trained several dedicated models tasked to extract important features from the raw input data as detailed in the Supplementary Information, Supplementary Notes 1 and 2.
We show that all those different embedding methods produce very close results (see Supplementary Figs\@. 1 and 2).
This means that the methodology presented here is robust to the particular choice of the embedding method to predict topological transitions.
While we used ResNet-50 v2 to demonstrate that an out-of-the-shelf method works well here, other models} could have been used instead, as discussed in the Supplementary Information.
Indeed, there is a variety of other models (such as ResNet-152~\cite{He2016}, VGG-16~\cite{simonyan2014very}, AlexNet~\cite{krizhevsky2014one}, and EfficientNetV2~\cite{tan2021efficientnetv2} to name a few) that have been trained on the same ImageNet dataset~\cite{Deng2009} as the ResNet model used in this study.
These models may prove to increase the sensitivity and accuracy of our results by enabling further clustering and disentangling of complex datasets in latent space.
%
%Another advantage \revision{of using pre-trained models} may be \revision{reduction in} the training time and the amount of data needed to achieve similar accuracy on the sensitivity score and therefore improve the selectivity of the present self-supervised learning approach.
%
As an alternative to neural network models, statistical and spectral-density functions~\cite{torquato2002statistical, niezgoda2013novel, bostanabad2018computational, montes2021accelerating} could be used for example as cost-effective, representative approaches to characterize microstructural patterns via a finite set of characteristic function(s) and/or feature(s).
As pointed out by Aggarwal \textit{et al}\@.~\cite{aggarwal2001surprising}, besides the improvements on these latent embedding approaches, the choice of a particular distance metric may significantly improve the results of these standard algorithms.
If the (microstructure) data lies on some smooth manifold and we are interested in preserving the local geometry of the microstructure phase-field data on that manifold, then using different distance metrics (\textit{i.e.} non Euclidean) when applying ResNet + PCA/UMAP to our data can shed alternative light on the same data and potentially further improve our classification of hierarchy of transition regimes.
}

\par{
In addition to the self-taught labelling of topological transitions, compared to other methods that focus solely on detecting transitions, our self-supervised approach can also provide additional insights into the role of the different input process parameters on the process-microstructure linkage.
By interpreting changes in sensitivity scores, we reveal the role of different input process parameters in controlling different pattern-formation regimes.
This includes which process parameters are responsible for the largest portion of the variation in microstructural patterns, and therefore which parameter(s) triggers specific classes of microstructural patterns.
For instance, our results reveal that the bulk mobility plays a role in controlling the complexity of the pattern seen in the physical vapor deposition problem.
Such linkage is important in the context of unraveling the mechanisms needed for fabricating multi-scale, \revision{microstructurally} precise materials.
}

\par{
In the results presented above, we show that predicting error is not only useful to determine pattern class transitions, but also for sensitivity analysis.
Namely, when the microstructure morphology changes slowly as a function of a given input process parameter, there will be a high error when mapping the final microstructural pattern back to that process parameter.
In contrast, when the morphology changes significantly, there is a narrow range where a process parameter reaches a particular microstructural pattern, and therefore error will be low.
This observation provides insight into the process parameters that need low or high tolerance.
In addition, some input process parameters may yield a class of microstructural consistently, while others may be affected by stochasticity in their initial values (see variability in sensitivity scores in Fig.~\ref{fig:pvd-inversemapping}a for instance).
This consistency metric will be inferred by the error when mapping control parameters to the microstructure morphology; in other words, much like a typical surrogate model, one can find the final microstructural pattern from initial conditions; error in this task will give insight into input process parameters that consistently, or inconsistently, \revision{yield} a given microstructure of interest.
}

\par{
Going forward, we note that our framework could be extended to other, more complex pattern-forming processes.
In the current work, we assessed topological transition in binary phase systems, but the same methods could also be applied to, for example, complex multi-component ternary or quaternary phase-separating processes~\cite{cogswell2011thermodynamic, kim2012phase}.
Further, our self-supervised-learning model did not have to be built exclusively on simulation-based data.
We chose to illustrate our concept via simulation to establish comprehensive data sets to serve as ground truths for model evaluation.
However, experimental microscopy data could have been used and combined with our simulation data to explore nuanced effects~\cite{weis1998morphological, hedstrom2012phase} on actual synthesis processes.
\revision{Another direction would be to increase the efficiency of our approach.
As discussed above, interpreting individual sensitivity score can be misleading.
Instead, one must average the individual scores on a previously prepared validation set.
However, it might be possible to directly measure the uncertainty of the network in solving the inverse problem. This could be done using \textit{e.g.}, the concrete dropout \cite{Gal2017}.
Consequently, it would allow us to measure the score on the individual sample-label, eliminating the need of averaging over a large number of samples.}
Although the approach used in this study could be improved both to enhance sensitivity and accuracy, this work is a step towards automatically detecting critical transitions during evolutionary systems and opens new ways for identifying discovering and understanding unseen or hard-to-\revision{discern} transition regimes in these systems.
}

%%%%%%%%%%%%%%%%%%%%%%%%%%%%%%%%%%%%%%%%%%%%%%%%%%%%%%%%%%%%%%%%%
%%%%%%%%%%%%%%%%%%%%%%%%%%%%%%%%%%%%%%%%%%%%%%%%%%%%%%%%%%%%%%%%%
%%
%% METHODS
%%
%%%%%%%%%%%%%%%%%%%%%%%%%%%%%%%%%%%%%%%%%%%%%%%%%%%%%%%%%%%%%%%%%
%%%%%%%%%%%%%%%%%%%%%%%%%%%%%%%%%%%%%%%%%%%%%%%%%%%%%%%%%%%%%%%%%   
\section*{METHODS}
%%
%% SPINODAL MODEL
%%
\subsection*{Spinodal decomposition of a two-phase mixture}
{
Spinodal decomposition of a two-phase mixture is one of the oldest and simplest phase-field problems~\cite{chen2002phase}.
This model looks into the diffusion of solutes within a matrix and serves as the basis for a large number of phase-field models.
In this model, a single compositional order parameter $c(\bf{x},t)$ is used to describe the atomic fraction of solute in space ($\bf{x}$) and time ($t$).

The evolution of $c$ is described via the Cahn-Hilliard equation~\cite{chen2002phase} such that,
\begin{equation}
    \frac{\partial c}{\partial t} = \nabla \cdot \left(M_{\rm{c}}(c)\nabla\left[\omega_{\rm{c}}(c^3-c) + \kappa_{\rm{c}}\nabla^2c\right]\right),
\end{equation}
where $\omega_{\rm{c}}$ is the energy-barrier height between the equilibrium phases and
$\kappa_c$ is the archetypal curvature-energy penalty term.
The concentration-dependent Cahn–Hilliard mobility is taken to be $M_{\rm{c}} = s(c)M_{\rm{A}} + (1-s(c))M_{\rm{B}}$, where $M_{\rm{i}}$ denotes the mobility of each phase.
The switching function $s(c)=\dfrac{1}{4}(2-c)(1+c)^2$ is a smooth interpolation function between the mobilities.
The values of the energy barrier height between the equilibrium phases and the gradient energy coefficients were assumed to be constant with $\omega_{\rm{c}}=\kappa_{\rm{c}}=1$.
\revision{The choice of the initial concentrations of phase A and B ($c_{\rm{A}}$ and $c_{\rm{B}}$) as well as the choice of mobilities for each phase ($M_{\rm{A}}$ and $M_{\rm{B}}$) constitute the input process parameter space and dictate the resulting microstructures.}
}
%%
%% PVD MODEL
%%
\subsection*{Physical vapor deposition of binary-alloy thin films}
{
The phase-field model used to simulate physical vapor deposition (PVD) is described in detail elsewhere~\cite{Stewart2020}.
This model uses conserved order-parameters, $\phi$ and $c$ to describe the structural and compositional ordering of the system as the thin film grows.
The field variable $\phi$ is meant to distinguishes between the vapor and solid phases and is used to track the growing thin film.
The compositional field variable $c$ distinguishes between the two phases of the growing binary alloy thin film.
The composition evolution is described via the Cahn-Hilliard equation such that,
\begin{equation}
    \frac{\partial c}{\partial t} = \nabla \cdot \left(M_c(\phi, c)\nabla\frac{\partial \mathcal{F}}{\partial t}\right),
\end{equation}
where $M_{\rm{c}}$ is the structurally and compositionally dependent mobility function and
$\mathcal{F}$ is the total free energy of the system.
The mobility functional $M_{\rm{c}}$ is constructed to describe both bulk and surface mobilities for each phase, $M^{\rm{Bulk}}_{\rm{i}}$ and $M^{\rm{Surf}}_{\rm{i}}$ respectively.
The complete expression for $\mathcal{F}$ is the same as in Stewart and Dingreville~\cite{Stewart2020} without the elasticity contributions.
\revision{The expression for $M_{\rm{c}}$ is provided in Eqs. (8) and (9) in Stewart and Dingreville~\cite{Stewart2020}.} 
To describe thin-film growth, a source term is incorporated in the Cahn–Hilliard equation, such that the evolution of the vapor-solid field variable, $\phi$, is given by,
\begin{equation}
    \frac{\partial \phi}{\partial t} = \nabla \cdot \left(M_{\rm{\phi}}(\phi)\nabla\frac{\partial \mathcal{F}}{\partial \phi}\right) + \nabla\phi\cdot(\rho{\bf v}),
\end{equation}
where $M_{\rm{\phi}}$ is the Cahn-Hilliard mobility.
The second term couples the thin-film evolution to the incident vapor flux (via a vapor density field $\rho$ and deposition rate $\bf{v}$) and acts as the source term for interfacial growth and surface roughening.
We defined a normalized deposition rate\cite{Lu2012} $v^{\rm{N}} = 2\sin\left(\alpha\right) |{\bf{v}}|/(M^{\rm{Bulk}}_{\rm{A}}+M^{\rm{Bulk}}_{\rm{B}})$ (where $\alpha$ is the deposition angle) which is meant to account for the competition between film growth and phase separation in the growing film.
\revision{In this problem, the choice of the initial concentrations of phase A and B ($c_{\rm{A}}$ and $c_{\rm{B}}$), the choice of bulk and surface mobilities for each phases ($M^{\rm{i}}_{\rm{A}}$ and $M^{\rm{i}}_{\rm{B}}$, with i being either Bulk or Surf), and the deposition rate $\bf{v}$ constitute the input process parameter space and dictate the resulting microstructures.}
}
%%
%% DATA PREPARATION
%%
\subsection*{Data preparation}
{
These two phase-field models have been implemented in Sandia’s in-house multi-physics phase-field modeling code: Mesoscale Multiphysics Phase-Field Simulator~\cite{Stewart2020, dingreville2020benchmark} (MEMPHIS).
\revision{Examples of simulation results for these two models are provided as supplementary information in Supplementary Note 3.}
The two-dimensional (2D), spatio-temporal evolution of the microstructural patterns (as captured through the $c$ compositional field variable in both models) were numerically solved using a finite-difference scheme with second-order central difference stencils for all spatial derivatives.
Numerical temporal integration of the equations was performed using the explicit Euler method.
All 2D simulations were performed on a uniform numerical grid of $512\times512$ for the spinodal decomposition model and $256\times256$ for the PVD model, a spatial discretization of $\Delta x = \Delta y = 1$, and a temporal discretization of $\Delta t = 10^{-4}$ for the spinodal decomposition model and a variable $\Delta t \in [10^{-3}, 10^{-2}]$ for the PVD model.
The composition field within the simulation domain was initially randomly populated by sampling a truncated Gaussian distribution between $-1$ and 1 with a standard deviation of 0.35.
Each spinodal decomposition simulation was run for 50,000,000 time steps with periodic boundary conditions applied to all sides of the domain.
The run time for the PVD simulations varied depending on the deposition rate to achieve a given film height.
}

{
We used a Latin Hypercube Sampling (LHS) approach to generate our training and testing datasets. For the spinodal decomposition model, we varied the phase concentrations and phase mobilities parameters. For the phase concentration parameter, we varied the concentration of phase A, $c_{\rm{A}} \in [0.15,0.85]$. Note that the phase concentration of B is simply $c_{\rm{B}}=1-c_{\rm{A}}$.
For the mobility parameters, we independently varied the mobility values over four orders of magnitude such that $M_{\rm{i}}\in[0.01, 100],~\rm{i=A~or~B}$.
For the PVD model, we keep the phase fraction at 50/50 and sampled the bulk and surface mobilities $M_{\rm{i}}^{\textrm{Bulk}} \in [0.1, 10]$, $M_{\rm{i}}^{\textrm{Surf}} \in [0.1, 100]$, as well as the deposition rate conditions $|\mathbf{v}| \in [0.1, 1]$.
For the spinodal decomposition model, this resulted in total 4998 simulations, 80\% of which was used to train the model (3998 instances) and the remaining 20\% of data (1000 instances) was used for testing.
For the PVD model, we selected the final microstructure pattern at the time step that corresponded to a film height of 224 pixels, resulting in a $224\times224$ image.
For this model, our dataset had 11775 simulations. We used 60\% of the data instances for training and 40\% for testing purposes.
}
%%
%% Network architecture selection
%%
\subsection*{\revision{Selection of neural network architecture}}
\revision{
 The outcome of detecting transitions is determined by both the initial condition (\textit{e.g.}, value of a random seed) and values of the process parameters (phase mobility, deposition rate in case of PVD, \textit{etc}.).
 While the local details of microstructural patterns depend on the specific initial conditions, the overall characteristics of those patterns are determined mostly by the values of the input process parameters.
 When solving the inverse problem, we wish to measure and gauge the role of the process parameters, not the random initialization of the simulations.
 We achieved this by first transforming the input using a CNN.
 This type of network is equivariant to translations of the input and therefore insensitive to relative shifts of the overall characteristics of the microstructural patterns that could be incurred by random initialization values of the simulations.
 We can use CNNs to extract features from the raw data, that are related to the overall shape of the patterns, not the specific position of those.
 See an extended discussion of that topic in Supplementary Information, Supplementary Note 3. 
 }
 
 \revision{
While there are many different possible convolutional architectures, one natural choice is to use a CNN from the ResNet family \cite{He2016a}.
ResNet networks are built with what are known as residual blocks.
CNNs are those with convolution layers, in which a small patch of an image is scanned, and features, such as colors and shades, of that small image are fed through a node.
There are multiple ways to weigh features in a patch, which are known as filters.
Residual blocks take features from earlier convolution layers and feed them wholesale into a layer that could be many levels away from the initial layer.
This allows for features to not get ``lost in translation'' when moving from one layer to the next.
This network can therefore achieve respectable predictive accuracy with comparatively more layers, thus capturing long-range and non-linear correlations between far-away parts of an image, or in this case, a microstructure evolution simulation.
 In effect, ResNet models can take advantage of deeper architectures reducing the impact of the infamous vanishing gradient problem \cite{Bengio1994}.
 We further elaborated on this topic in Supplementary Information, Supplementary Note 4.
}
%%
%% LATENT EMBEDDING
%%
\subsection*{Latent embedding}
{
We embedded microstructural pattern images using ResNet-50 v2 \cite{He2016}, a popular model that has been used to automatically classify a variety of images.
ResNet-50 v2 is a fifty-layer-deep CNN.
In particular, the ResNet-50 v2 model used in this work was trained on \revision{images from the ImageNet~\cite{Deng2009} database but not on the phase-field data.
Training of the ResNet-50 v2 model was first done by He \textit{et al}\@.\cite{He2016}.
The model trained on ImageNet learns to extract a hierarchy of features, from simple geometric shapes (vertical, horizontal, or skew lines, shapes of different sizes, lines of different curvatures, \textit{etc}.) to more complex, composed patterns.
Images present in the ImageNet database depict 1000 classes such as different types of animals, everyday objects, and various building structures.
While most of the training examples represent colorful (3-channel) images, there is also a small proportion (about $\sim$2\%) of monochromatic images.
To make the phase-field simulation data input compatible with our model, we copied each image three times (created 3 identical pseudo-color channels).
While it creates some redundancy (the network might detect ``red'', ``green'' and ``blue'' edges of the same orientation), it did not affect the ability to produce a consistent embedding for the collection of our data.
Examples and details of the data used for training the ResNet-50 v2 model are presented in the Supplementary Information, Supplementary Note 4.
}
We took all but the classification layer of this pre-trained model to embed simulation images into several hundred dimensional latent feature space.
The size of this feature space is small enough to reduce the feature dimension from 65536 ($256 \times 256$ pixels where we treat each pixel as a feature) to something that is far more manageable, 2048 latent dimensions (\textit{i.e.} 1\% of this), but this space is large enough to capture nuances in microstructural patterns from subtle variations in control parameters or initial conditions.
Parameter predictions are then made by applying a few dense all-to-all connection neural network layers to this set of latent features.
}
%%
%% INVERSE MAPPING
%%
\subsection*{Inverse mapping workflow}
{
{\textit{Spinodal decomposition}:}
The original simulations were saved as $4998 \times 101 \times 512 \times 512$ (simulation number, time frame, X position, Y position) dimensional numpy arrays, with \texttt{float16} numbers that range from around $-1$ to $+1$.
We reduced the dimensions of each frame from $512\times512$ to $256\times256$ (replacing every block of $2\times2$ pixels by the mean value).
From the $4998$ original simulations, we randomly selected 1000 of them ($20\%$) to form a validation set.
We used the remaining $3998$ simulations to construct the training set.
%
%To create the feature vector from an image, we can use one of the pre-trained models available in TensorFlow Hub.
%
%Here, we decided to use the ResNet-50 v2 model.
%
%The key difference with the V1 is the use of batch normalization.
%
%ResNet-50 has 50 layers and was trained on standard ImageNet-1k dataset (ILSVRC-2012-CLS).
%
%The feature latent vector has a size of 2048.
%
The input images are expected to have color values in the range [0,1] and the expected size of the input images is $224\times224$ pixels by default, but other input sizes are possible.
Because the ResNet-50 v2 model requires a 3-channel input, we just tripled our output (each channel has the same pictures).
The inverse problem was solved by training a neural network, a simple feedforward neural network with two hidden layers (with $512$ and $1024$ neurons, respectively).
The output of the network was a dense layer with two neurons with linear activation -- one neuron to predict mobility of the phase A, and another to predict mobility of the phase B.
We used dropout layers and ReLU activation functions.
Schematically, the network architecture was $\mbox{\texttt{Dense}}_{512}^{\mbox{\scriptsize ReLU}} \times \mbox{\texttt{Dropout}}_{0.25} \times \mbox{\texttt{Dense}}_{1024}^{\mbox{\scriptsize ReLU}} \times \mbox{\texttt{Dropout}}_{0.25} \times \mbox{\texttt{Dense}}_{2}^{\mbox{\scriptsize linear}}$.
We trained the model to predict the correct values of mobility, with the mean squared error (MSE) loss function. We applied early stopping with the patience $4$ and tolerance of $0.0001$.
}

{
{\textit{Physical vapor deposition}:}
The original simulations were saved as $11775\times 51 \times 256\times256$ (simulation number, time frame, X position, Y position) dimensional numpy arrays.
For each simulation, we selected the time step that corresponds to a thin film height of 224 pixels to avoid capturing any surface roughness effects.
We rejected simulations that had heights smaller than 224 (this simulation corresponded to extremely slow deposition rates; only 5 simulations were rejected in the process).
We cut four $224\times224$ squares from each original frame, translating the cut origin by 64 pixels each time (we apply horizontal periodic boundary conditions). This gave us $11770\cdot4=47080$ individual data points.
We truncated values larger than 1, and smaller than -1, and then we scaled everything from 0 to 1 (we applied transformation $x \rightarrow (x+1)/2$).
To speed-up the next step, we reduced the dimension of each square by two, from $224 \times 224$ to $112 \times 112$ pixels.
We used the pre-trained ResNet-50 v2 model to produce 2048-dimension latent embeddings of each square (to do this, we had to emulate the 3 channel colors by copied the one-channel input three times, just as we did for spinodal decomposition).
Since original simulations were repeated 5 times for each unique set of parameters, the $47080$ data points correspond to 2354 unique sets of initial parameters.
We reserved $40\%$ of the data for the validation set (18840 input examples, corresponding to 942 unique sets of initial parameters).
The remaining $60\%$ was used for training (28240 input examples, corresponding to 1412 unique sets of initial parameters).
The inverse problem was solved by training a neural network, a simple feed-forward neural network with two  hidden layer (similar to the architecture used for spinodal decomposition), namely
$\mbox{\texttt{Dense}}_{512}^{\mbox{\scriptsize ReLU}} \times \mbox{\texttt{Dropout}}_{0.25} \times \mbox{\texttt{Dense}}_{1024}^{\mbox{\scriptsize ReLU}} \times \mbox{\texttt{Dropout}}_{0.25} \times \mbox{\texttt{Dense}}_{1}^{\mbox{\scriptsize linear}}$.
The only modification \revision{with respect to the previous setup} was, that we used here only a single output neuron. 
The motivation for this was that was that the predicted parameters had different physical units. When predicting multiple values at once, the MSE loss function would be sensitive to the particular choice of the units for each parameter. To address this issue, we decided to train separate network for each control parameter -- thus, the single output unit in the architecture above.
We trained the network with early stopping with the patience $4$ and tolerance of $0.001$.
}
%%
%% ResNet Training and Validation
%%
   
\subsection*{\revision{Model training setup for the inverse problem}}
\revision{ 
 What matters when solving the inverse problem is the relative performance.
 This situation is similar to that described in the ``Learning phase transitions by confusion'' paper by van Nieuwenburg \textit{et al}\@.\cite{Nieuwenburg2017}.
 In that work, the proposition was to use a small feedforward neural network, with one hidden layer and 80 neurons.
 There are two possible regimes where neural networks generalize well.
 One is when the number of training examples is larger than the number of trainable parameters.
 Second, when the number of trainable parameters is much larger than the number of training examples (an overparameterized  region) \cite{dAscoli2020}.
 Keeping in mind that in practical applications the number of training examples can be limited, we decided to explore the second regime.
 As discussed in the next section, to solve the inverse problem we used a network with two hidden layers, having 512 and 1024 neurons, respectively. 
 %To prevent overfitting, we applied dropout, data augmentation and early stop.
 We further elaborated on this topic in Supplementary Information, Supplementary Note 5.
}

%%
%% Inverse Problem Training and Validation
%%
   
\subsection*{\revision{Training and validation of the inverse problem network}}
\begin{figure}[h!]
        \centering
        \includegraphics[width=0.7\textwidth]{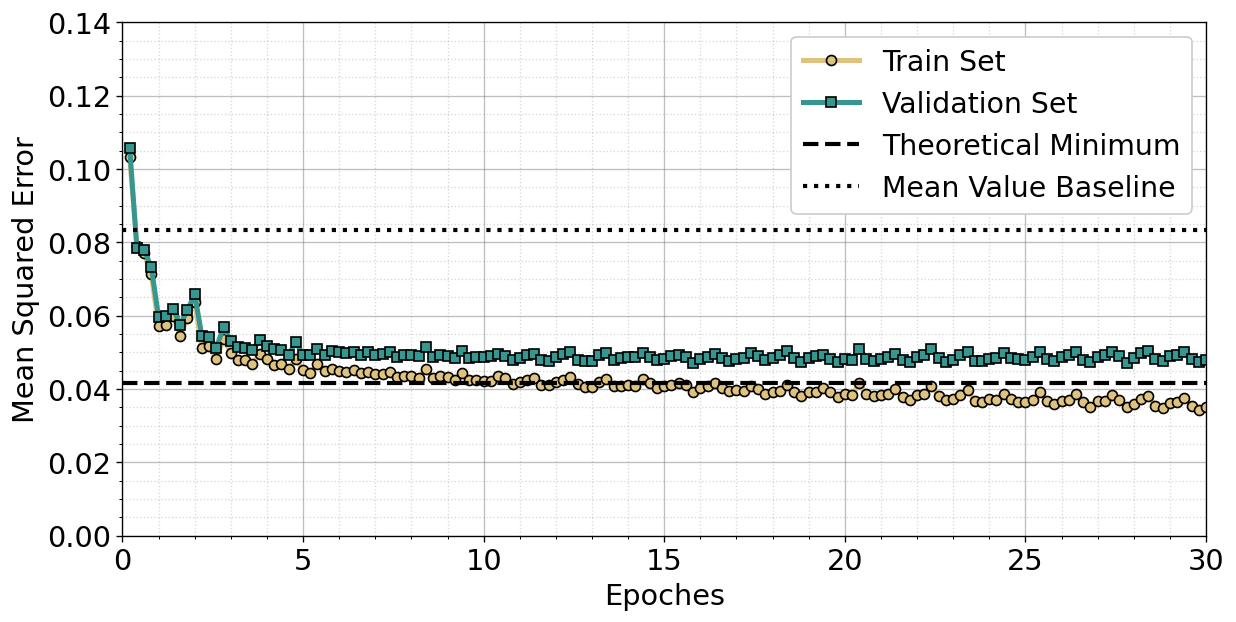}
        \caption{\revision{{\bf Training logs for the feedforward network solving the inverse problem.} The mean value baseline corresponds to $\Delta_{\rm A}$ and the theoretical minimum to $\bar\Delta$.
        Here, we used a network with two hidden layers, having 1024 and 512 neurons, respectively, $\mbox{\texttt{Dense}}_{512}^{\mbox{\scriptsize ReLU}} \times \mbox{\texttt{Dropout}}_{0.25} \times \mbox{\texttt{Dense}}_{1024}^{\mbox{\scriptsize ReLU}} \times \mbox{\texttt{Dropout}}_{0.25} \times \mbox{\texttt{Dense}}_{2}^{\mbox{\scriptsize linear}}$.}}
        \label{fig:training_log}
    \end{figure}

\revision{ 
To minimize any potential overfitting, we followed best practices during the training.
Namely, we used a validation set to track the generalization error, we used dropout to combat the overfitting, we used data augmentation (we used the advantage of the periodic boundaries conditions and we prepared several cropped versions of the input from each sample), and when training we used an early stop.
We visualize the training and validation error in case of spinodal decomposition data in Fig.~\ref{fig:training_log}.
In case of the spinodal decomposition, we can easily determine the theoretical minimum for the mean squared error.
The outcome of the process is determined by the value of mobility of the majority phase.
Thus, assuming uniform distribution of samples and normalization of the target values to [0, 1], the best we can hope is to have zero error, $\Delta_{\rm B}$, when predicting mobility of the majority phase and an error of $\Delta_{\rm A} = \int_0^1 (x-0.5)^2\, dx = 1/12$ when predicting the mobility value of the minority phase.
Consequently, the lower bound for the cumulative mean square error is $ \bar\Delta = (\Delta_{\rm A} + \Delta_{\rm B})/2 = \left(1/12 + 0\right)/2 \approx 0.0416(6)$. 
Looking at Fig\@.~\ref{fig:training_log}, we see that the validation set approaches the theoretical limit of the irreducible/intrinsic error, indicating a relatively good generalization capability and good quality of training.
Additional details on the training performance of the ResNet-50 model and comparison with other pre-trained models are provided in the Supplementary Information in Supplementary Note 5.
}

%%%%%%%%%%%%%%%%%%%%%%%%%%%%%%%%%%%%%%%%%%%%%%%%%%%%%%%%%%%%%%%%%
%%%%%%%%%%%%%%%%%%%%%%%%%%%%%%%%%%%%%%%%%%%%%%%%%%%%%%%%%%%%%%%%%
%%
%% CODE/DATA AVAILABILITY
%%
%%%%%%%%%%%%%%%%%%%%%%%%%%%%%%%%%%%%%%%%%%%%%%%%%%%%%%%%%%%%%%%%%
%%%%%%%%%%%%%%%%%%%%%%%%%%%%%%%%%%%%%%%%%%%%%%%%%%%%%%%%%%%%%%%%%

\section*{DATA AVAILABILITY}

Samples of our datasets are available at \url{https://github.com/marcinabram/topological_transitions/tree/main/data_sample}.
The full training data sets as well as all validation and test cases are available from the corresponding author upon reasonable request.

%Data will be made available upon acceptation of this manuscript. This statement will be updated at publication.

%%%%%%%%%%%%%%%%%%%%%%%%%%%%%%%%%%%%%%%%%%%%%%%%%%%%%%%%%%%%%%%%%:~

\section*{CODE AVAILABILITY}

Source code for training and evaluating the machine learning models is available at \url{https://github.com/marcinabram/topological_transitions}.

%Code will be made available upon acceptation of this manuscript. This statement will be updated at publication.

%%%%%%%%%%%%%%%%%%%%%%%%%%%%%%%%%%%%%%%%%%%%%%%%%%%%%%%%%%%%%%%%%
%%%%%%%%%%%%%%%%%%%%%%%%%%%%%%%%%%%%%%%%%%%%%%%%%%%%%%%%%%%%%%%%%
%%
%% REFERENCES
%%
%%%%%%%%%%%%%%%%%%%%%%%%%%%%%%%%%%%%%%%%%%%%%%%%%%%%%%%%%%%%%%%%%
%%%%%%%%%%%%%%%%%%%%%%%%%%%%%%%%%%%%%%%%%%%%%%%%%%%%%%%%%%%%%%%%%
\bibliographystyle{naturemag}  % Nature Bibtex Style
\def\bibsection{\section*{\refname}}
\bibliography{main-revision.bib}

\begin{thebibliography}{10}
\expandafter\ifx\csname url\endcsname\relax
  \def\url#1{\texttt{#1}}\fi
\expandafter\ifx\csname urlprefix\endcsname\relax\def\urlprefix{URL }\fi
\providecommand{\bibinfo}[2]{#2}
\providecommand{\eprint}[2][]{\url{#2}}

\bibitem{viamontes2006isotropic}
\bibinfo{author}{Viamontes, J.}, \bibinfo{author}{Oakes, P.~W.} \&
  \bibinfo{author}{Tang, J.~X.}
\newblock \bibinfo{title}{Isotropic to nematic liquid crystalline phase
  transition of {\textit f}-actin varies from continuous to first order}.
\newblock \emph{\bibinfo{journal}{Phys. Rev. Lett.}}
  \textbf{\bibinfo{volume}{97}}, \bibinfo{pages}{118103}
  (\bibinfo{year}{2006}).
\newblock \urlprefix\url{https://doi.org/10.1103/PhysRevLett.97.118103}.

\bibitem{antal1999formation}
\bibinfo{author}{Antal, T.}, \bibinfo{author}{Droz, M.},
  \bibinfo{author}{Magnin, J.} \& \bibinfo{author}{R{\'a}cz, Z.}
\newblock \bibinfo{title}{Formation of {L}iesegang patterns: A spinodal
  decomposition scenario}.
\newblock \emph{\bibinfo{journal}{Phys. Rev. Lett.}}
  \textbf{\bibinfo{volume}{83}}, \bibinfo{pages}{2880} (\bibinfo{year}{1999}).
\newblock \urlprefix\url{https://doi.org/10.1103/PhysRevLett.83.2880}.

\bibitem{toramaru2003experimental}
\bibinfo{author}{Toramaru, A.}, \bibinfo{author}{Harada, T.} \&
  \bibinfo{author}{Okamura, T.}
\newblock \bibinfo{title}{Experimental pattern transitions in a {L}iesegang
  system}.
\newblock \emph{\bibinfo{journal}{Phys. D: Nonlinear Phenom.}}
  \textbf{\bibinfo{volume}{183}}, \bibinfo{pages}{133--140}
  (\bibinfo{year}{2003}).
\newblock \urlprefix\url{https://doi.org/10.1016/S0167-2789(03)00139-8}.

\bibitem{shimizu2017liesegang}
\bibinfo{author}{Shimizu, Y.}, \bibinfo{author}{Matsui, J.},
  \bibinfo{author}{Unoura, K.} \& \bibinfo{author}{Nabika, H.}
\newblock \bibinfo{title}{Liesegang mechanism with a gradual phase transition}.
\newblock \emph{\bibinfo{journal}{J. Phys. Chem. B}}
  \textbf{\bibinfo{volume}{121}}, \bibinfo{pages}{2495--2501}
  (\bibinfo{year}{2017}).
\newblock \urlprefix\url{https://doi.org/10.1021/acs.jpcb.7b01275}.

\bibitem{nabika2019pattern}
\bibinfo{author}{Nabika, H.}, \bibinfo{author}{Itatani, M.} \&
  \bibinfo{author}{Lagzi, I.}
\newblock \bibinfo{title}{Pattern formation in precipitation reactions: The
  {L}iesegang phenomenon}.
\newblock \emph{\bibinfo{journal}{Langmuir}} \textbf{\bibinfo{volume}{36}},
  \bibinfo{pages}{481--497} (\bibinfo{year}{2019}).
\newblock \urlprefix\url{https://doi.org/10.1021/acs.langmuir.9b03018}.

\bibitem{sakurai1993morphology}
\bibinfo{author}{Sakurai, S.} \emph{et~al.}
\newblock \bibinfo{title}{Morphology transition from cylindrical to lamellar
  microdomains of block copolymers}.
\newblock \emph{\bibinfo{journal}{Macromolecules}}
  \textbf{\bibinfo{volume}{26}}, \bibinfo{pages}{485--491}
  (\bibinfo{year}{1993}).
\newblock \urlprefix\url{https://doi.org/10.1021/ma00055a013}.

\bibitem{castelletto2004morphologies}
\bibinfo{author}{Castelletto, V.} \& \bibinfo{author}{Hamley, I.~W.}
\newblock \bibinfo{title}{Morphologies of block copolymer melts}.
\newblock \emph{\bibinfo{journal}{Curr. Opin. Solid State Mater. Sci.}}
  \textbf{\bibinfo{volume}{8}}, \bibinfo{pages}{426--438}
  (\bibinfo{year}{2004}).
\newblock \urlprefix\url{https://doi.org/10.1016/j.cossms.2005.06.001}.

\bibitem{Lu2012}
\bibinfo{author}{Lu, Y.} \emph{et~al.}
\newblock \bibinfo{title}{Microstructure map for self-organized phase
  separation during film deposition}.
\newblock \emph{\bibinfo{journal}{Phys. Rev. Lett.}}
  \textbf{\bibinfo{volume}{109}} (\bibinfo{year}{2012}).
\newblock \urlprefix\url{https://doi.org/10.1103/physrevlett.109.086101}.

\bibitem{herman2020data}
\bibinfo{author}{Herman, E.}, \bibinfo{author}{Stewart, J.~A.} \&
  \bibinfo{author}{Dingreville, R.}
\newblock \bibinfo{title}{A data-driven surrogate model to rapidly predict
  microstructure morphology during physical vapor deposition}.
\newblock \emph{\bibinfo{journal}{Appl. Math. Model.}}
  \textbf{\bibinfo{volume}{88}}, \bibinfo{pages}{589--603}
  (\bibinfo{year}{2020}).
\newblock \urlprefix\url{https://doi.org/10.1016/j.apm.2020.06.046}.

\bibitem{powers2020microstructural}
\bibinfo{author}{Powers, M.}, \bibinfo{author}{Derby, B.},
  \bibinfo{author}{Shaw, A.}, \bibinfo{author}{Raeker, E.} \&
  \bibinfo{author}{Misra, A.}
\newblock \bibinfo{title}{Microstructural characterization of phase-separated
  co-deposited {Cu}--{Ta} immiscible alloy thin films}.
\newblock \emph{\bibinfo{journal}{J. Mater. Res.}}
  \textbf{\bibinfo{volume}{35}}, \bibinfo{pages}{1531--1542}
  (\bibinfo{year}{2020}).
\newblock \urlprefix\url{https://doi.org/10.1557/jmr.2020.100}.

\bibitem{powers2021compositionally}
\bibinfo{author}{Powers, M.}, \bibinfo{author}{Stewart, J.~A.},
  \bibinfo{author}{Dingreville, R.}, \bibinfo{author}{Derby, B.~K.} \&
  \bibinfo{author}{Misra, A.}
\newblock \bibinfo{title}{Compositionally-driven formation mechanism of
  hierarchical morphologies in co-deposited immiscible alloy thin films}.
\newblock \emph{\bibinfo{journal}{Nanomaterials}}
  \textbf{\bibinfo{volume}{11}}, \bibinfo{pages}{2635} (\bibinfo{year}{2021}).
\newblock \urlprefix\url{https://doi.org/10.3390/nano11102635}.

\bibitem{landau1937theorie}
\bibinfo{author}{Landau, L.~D.}
\newblock \bibinfo{title}{Zur theorie der phasenumwandlungen {II}}.
\newblock \emph{\bibinfo{journal}{Phys. Z. Sowjetunion}}
  \textbf{\bibinfo{volume}{11}}, \bibinfo{pages}{26--35}
  (\bibinfo{year}{1937}).

\bibitem{muller1999variational}
\bibinfo{author}{M{\"u}ller, S.}
\newblock \bibinfo{title}{Variational models for microstructure and phase
  transitions}.
\newblock In \emph{\bibinfo{booktitle}{Calculus of Variations and Geometric
  Evolution Problems}}, \bibinfo{pages}{85--210} (\bibinfo{publisher}{Springer,
  Berlin}, \bibinfo{year}{1999}).
\newblock \urlprefix\url{https://doi.org/10.1007/BFb0092670}.

\bibitem{kosterlitz1973ordering}
\bibinfo{author}{Kosterlitz, J.~M.} \& \bibinfo{author}{Thouless, D.~J.}
\newblock \bibinfo{title}{Ordering, metastability and phase transitions in
  two-dimensional systems}.
\newblock \emph{\bibinfo{journal}{J. Phys. C Solid State Phys.}}
  \textbf{\bibinfo{volume}{6}}, \bibinfo{pages}{1181} (\bibinfo{year}{1973}).
\newblock \urlprefix\url{https://doi.org/10.1088/0022-3719/6/7/010}.

\bibitem{bagchi1996computer}
\bibinfo{author}{Bagchi, K.}, \bibinfo{author}{Andersen, H.~C.} \&
  \bibinfo{author}{Swope, W.}
\newblock \bibinfo{title}{Computer simulation study of the melting transition
  in two dimensions}.
\newblock \emph{\bibinfo{journal}{Phys. Rev. Lett.}}
  \textbf{\bibinfo{volume}{76}}, \bibinfo{pages}{255} (\bibinfo{year}{1996}).
\newblock \urlprefix\url{https://doi.org/10.1103/PhysRevLett.76.255}.

\bibitem{bel2021geometrical}
\bibinfo{author}{Bel-Hadj-Aissa, G.}, \bibinfo{author}{Gori, M.},
  \bibinfo{author}{Franzosi, R.} \& \bibinfo{author}{Pettini, M.}
\newblock \bibinfo{title}{Geometrical and topological study of the
  {Kosterlitz--Thouless} phase transition in the xy model in two dimensions}.
\newblock \emph{\bibinfo{journal}{J. Stat. Mech. Theory Exp.}}
  \textbf{\bibinfo{volume}{2021}}, \bibinfo{pages}{023206}
  (\bibinfo{year}{2021}).
\newblock \urlprefix\url{https://doi.org/10.1088/1742-5468/abda27}.

\bibitem{Stewart2020}
\bibinfo{author}{Stewart, J.~A.} \& \bibinfo{author}{Dingreville, R.}
\newblock \bibinfo{title}{Microstructure morphology and concentration
  modulation of nanocomposite thin-films during simulated physical vapor
  deposition}.
\newblock \emph{\bibinfo{journal}{Acta Mater.}} \textbf{\bibinfo{volume}{188}},
  \bibinfo{pages}{181--191} (\bibinfo{year}{2020}).
\newblock \urlprefix\url{https://doi.org/10.1016/j.actamat.2020.02.011}.

\bibitem{carrasquilla2017machine}
\bibinfo{author}{Carrasquilla, J.} \& \bibinfo{author}{Melko, R.~G.}
\newblock \bibinfo{title}{Machine learning phases of matter}.
\newblock \emph{\bibinfo{journal}{Nat. Phys.}} \textbf{\bibinfo{volume}{13}},
  \bibinfo{pages}{431--434} (\bibinfo{year}{2017}).
\newblock \urlprefix\url{https://doi.org/10.1038/nphys4035}.

\bibitem{wei2017identifying}
\bibinfo{author}{Wei, Q.}, \bibinfo{author}{Melko, R.~G.} \&
  \bibinfo{author}{Chen, J.~Z.}
\newblock \bibinfo{title}{Identifying polymer states by machine learning}.
\newblock \emph{\bibinfo{journal}{Phys. Rev. E}} \textbf{\bibinfo{volume}{95}},
  \bibinfo{pages}{032504} (\bibinfo{year}{2017}).
\newblock \urlprefix\url{https://doi.org/10.1103/PhysRevE.95.032504}.

\bibitem{li2018applications}
\bibinfo{author}{Li, C.-D.}, \bibinfo{author}{Tan, D.-R.} \&
  \bibinfo{author}{Jiang, F.-J.}
\newblock \bibinfo{title}{Applications of neural networks to the studies of
  phase transitions of two-dimensional potts models}.
\newblock \emph{\bibinfo{journal}{Ann. Phys.}} \textbf{\bibinfo{volume}{391}},
  \bibinfo{pages}{312--331} (\bibinfo{year}{2018}).
\newblock \urlprefix\url{https://doi.org/10.1016/j.aop.2018.02.018}.

\bibitem{casert2019interpretable}
\bibinfo{author}{Casert, C.}, \bibinfo{author}{Vieijra, T.},
  \bibinfo{author}{Nys, J.} \& \bibinfo{author}{Ryckebusch, J.}
\newblock \bibinfo{title}{Interpretable machine learning for inferring the
  phase boundaries in a nonequilibrium system}.
\newblock \emph{\bibinfo{journal}{Phys. Rev. E}} \textbf{\bibinfo{volume}{99}},
  \bibinfo{pages}{023304} (\bibinfo{year}{2019}).
\newblock \urlprefix\url{https://doi.org/10.1103/PhysRevE.99.023304}.

\bibitem{zhang2019machine}
\bibinfo{author}{Zhang, W.}, \bibinfo{author}{Liu, J.} \& \bibinfo{author}{Wei,
  T.-C.}
\newblock \bibinfo{title}{Machine learning of phase transitions in the
  percolation and {XY} models}.
\newblock \emph{\bibinfo{journal}{Phys. Rev. E}} \textbf{\bibinfo{volume}{99}},
  \bibinfo{pages}{032142} (\bibinfo{year}{2019}).
\newblock \urlprefix\url{https://doi.org/10.1103/PhysRevE.99.032142}.

\bibitem{wang2016discovering}
\bibinfo{author}{Wang, L.}
\newblock \bibinfo{title}{Discovering phase transitions with unsupervised
  learning}.
\newblock \emph{\bibinfo{journal}{Phys. Rev. B}} \textbf{\bibinfo{volume}{94}},
  \bibinfo{pages}{195105} (\bibinfo{year}{2016}).
\newblock \urlprefix\url{https://doi.org/10.1103/PhysRevB.94.195105}.

\bibitem{hu2017discovering}
\bibinfo{author}{Hu, W.}, \bibinfo{author}{Singh, R.~R.} \&
  \bibinfo{author}{Scalettar, R.~T.}
\newblock \bibinfo{title}{Discovering phases, phase transitions, and crossovers
  through unsupervised machine learning: A critical examination}.
\newblock \emph{\bibinfo{journal}{Phys. Rev. E}} \textbf{\bibinfo{volume}{95}},
  \bibinfo{pages}{062122} (\bibinfo{year}{2017}).
\newblock \urlprefix\url{https://doi.org/10.1103/PhysRevE.95.062122}.

\bibitem{Nieuwenburg2017}
\bibinfo{author}{van Nieuwenburg, E.~P.~L.}, \bibinfo{author}{Liu, Y.-H.} \&
  \bibinfo{author}{Huber, S.~D.}
\newblock \bibinfo{title}{Learning phase transitions by confusion}.
\newblock \emph{\bibinfo{journal}{Nat. Phys.}} \textbf{\bibinfo{volume}{13}},
  \bibinfo{pages}{435--439} (\bibinfo{year}{2017}).
\newblock \urlprefix\url{https://doi.org/10.1038/nphys4037}.

\bibitem{wetzel2017unsupervised}
\bibinfo{author}{Wetzel, S.~J.}
\newblock \bibinfo{title}{Unsupervised learning of phase transitions: From
  principal component analysis to variational autoencoders}.
\newblock \emph{\bibinfo{journal}{Phys. Rev. E}} \textbf{\bibinfo{volume}{96}},
  \bibinfo{pages}{022140} (\bibinfo{year}{2017}).
\newblock \urlprefix\url{https://doi.org/10.1103/PhysRevE.96.022140}.

\bibitem{liu2018discriminative}
\bibinfo{author}{Liu, Y.-H.} \& \bibinfo{author}{Van~Nieuwenburg, E.~P.}
\newblock \bibinfo{title}{Discriminative cooperative networks for detecting
  phase transitions}.
\newblock \emph{\bibinfo{journal}{Phys. Rev. Lett.}}
  \textbf{\bibinfo{volume}{120}}, \bibinfo{pages}{176401}
  (\bibinfo{year}{2018}).
\newblock \urlprefix\url{https://doi.org/10.1103/PhysRevLett.120.176401}.

\bibitem{yoshioka2018learning}
\bibinfo{author}{Yoshioka, N.}, \bibinfo{author}{Akagi, Y.} \&
  \bibinfo{author}{Katsura, H.}
\newblock \bibinfo{title}{Learning disordered topological phases by statistical
  recovery of symmetry}.
\newblock \emph{\bibinfo{journal}{Phys. Rev. B}} \textbf{\bibinfo{volume}{97}},
  \bibinfo{pages}{205110} (\bibinfo{year}{2018}).
\newblock \urlprefix\url{https://doi.org/10.1103/PhysRevB.97.205110}.

\bibitem{rodriguez2019identifying}
\bibinfo{author}{Rodriguez-Nieva, J.~F.} \& \bibinfo{author}{Scheurer, M.~S.}
\newblock \bibinfo{title}{Identifying topological order through unsupervised
  machine learning}.
\newblock \emph{\bibinfo{journal}{Nat. Phys.}} \textbf{\bibinfo{volume}{15}},
  \bibinfo{pages}{790--795} (\bibinfo{year}{2019}).
\newblock \urlprefix\url{https://doi.org/10.1038/s41567-019-0512-x}.

\bibitem{lee2022phase}
\bibinfo{author}{Lee, K.}, \bibinfo{author}{Ayyasamy, M.},
  \bibinfo{author}{Delsa, P.}, \bibinfo{author}{Hartness, T.} \&
  \bibinfo{author}{Balachandran, P.}
\newblock \bibinfo{title}{Phase classification of multi-principal element
  alloys via interpretable machine learning}.
\newblock \emph{\bibinfo{journal}{npj Comput. Mater.}}
  \textbf{\bibinfo{volume}{8}} (\bibinfo{year}{2022}).
\newblock \urlprefix\url{https://doi.org/10.1038/s41524-022-00704-y}.

\bibitem{doersch2017multi}
\bibinfo{author}{Doersch, C.} \& \bibinfo{author}{Zisserman, A.}
\newblock \bibinfo{title}{Multi-task self-supervised visual learning}.
\newblock In \emph{\bibinfo{booktitle}{{IEEE} International Conference on
  Computer Vision, {ICCV} 2017, Venice, Italy, October 22-29, 2017}},
  \bibinfo{pages}{2070--2079} (\bibinfo{publisher}{{IEEE} Computer Society},
  \bibinfo{year}{2017}).
\newblock \urlprefix\url{https://doi.org/10.1109/ICCV.2017.226}.

\bibitem{zbontar2021barlow}
\bibinfo{author}{Zbontar, J.}, \bibinfo{author}{Jing, L.},
  \bibinfo{author}{Misra, I.}, \bibinfo{author}{LeCun, Y.} \&
  \bibinfo{author}{Deny, S.}
\newblock \bibinfo{title}{Barlow twins: Self-supervised learning via redundancy
  reduction}.
\newblock In \bibinfo{editor}{Meila, M.} \& \bibinfo{editor}{Zhang, T.} (eds.)
  \emph{\bibinfo{booktitle}{Proceedings of the 38th International Conference on
  Machine Learning}}, vol. \bibinfo{volume}{139} of
  \emph{\bibinfo{series}{Proceedings of Machine Learning Research}},
  \bibinfo{pages}{12310--12320} (\bibinfo{publisher}{PMLR},
  \bibinfo{year}{2021}).
\newblock \urlprefix\url{https://proceedings.mlr.press/v139/zbontar21a.html}.

\bibitem{chen2020simple}
\bibinfo{author}{Chen, T.}, \bibinfo{author}{Kornblith, S.},
  \bibinfo{author}{Norouzi, M.} \& \bibinfo{author}{Hinton, G.}
\newblock \bibinfo{title}{A simple framework for contrastive learning of visual
  representations}.
\newblock In \bibinfo{editor}{III, H.~D.} \& \bibinfo{editor}{Singh, A.} (eds.)
  \emph{\bibinfo{booktitle}{Proceedings of the 37th International Conference on
  Machine Learning}}, vol. \bibinfo{volume}{119} of
  \emph{\bibinfo{series}{Proceedings of Machine Learning Research}},
  \bibinfo{pages}{1597--1607} (\bibinfo{publisher}{PMLR},
  \bibinfo{year}{2020}).
\newblock \urlprefix\url{https://proceedings.mlr.press/v119/chen20j.html}.

\bibitem{He2016}
\bibinfo{author}{He, K.}, \bibinfo{author}{Zhang, X.}, \bibinfo{author}{Ren,
  S.} \& \bibinfo{author}{Sun, J.}
\newblock \bibinfo{title}{Identity mappings in deep residual networks}.
\newblock In \emph{\bibinfo{booktitle}{Computer Vision {\textendash} {ECCV}
  2016}}, \bibinfo{pages}{630--645} (\bibinfo{publisher}{Springer International
  Publishing}, \bibinfo{year}{2016}).
\newblock \urlprefix\url{https://doi.org/10.1007/978-3-319-46493-0_38}.

\bibitem{lee2019confusion}
\bibinfo{author}{Lee, S.~S.} \& \bibinfo{author}{Kim, B.~J.}
\newblock \bibinfo{title}{Confusion scheme in machine learning detects double
  phase transitions and quasi-long-range order}.
\newblock \emph{\bibinfo{journal}{Phys. Rev. E}} \textbf{\bibinfo{volume}{99}},
  \bibinfo{pages}{043308} (\bibinfo{year}{2019}).
\newblock \urlprefix\url{https://doi.org/10.1103/PhysRevE.99.043308}.

\bibitem{abdi2010principal}
\bibinfo{author}{Abdi, H.} \& \bibinfo{author}{Williams, L.~J.}
\newblock \bibinfo{title}{Principal component analysis}.
\newblock \emph{\bibinfo{journal}{Wiley Interdiscip. Rev. Comput. Stat.}}
  \textbf{\bibinfo{volume}{2}}, \bibinfo{pages}{433--459}
  (\bibinfo{year}{2010}).
\newblock \urlprefix\url{https://doi.org/10.1002/wics.101}.

\bibitem{mcinnes2018umap}
\bibinfo{author}{McInnes, L.}, \bibinfo{author}{Healy, J.} \&
  \bibinfo{author}{Melville, J.}
\newblock \bibinfo{title}{Umap: Uniform manifold approximation and projection
  for dimension reduction}.
\newblock \emph{\bibinfo{journal}{arXiv preprint arXiv:1802.03426}}
  (\bibinfo{year}{2018}).
\newblock \urlprefix\url{https://doi.org/10.48550/arXiv.1802.03426}.

\bibitem{McInnes_2018b}
\bibinfo{author}{McInnes, L.}, \bibinfo{author}{Healy, J.},
  \bibinfo{author}{Saul, N.} \& \bibinfo{author}{Grossberger, L.}
\newblock \bibinfo{title}{{UMAP: Uniform Manifold Approximation and
  Projection}}.
\newblock \emph{\bibinfo{journal}{J. Open Source Softw.}}
  \textbf{\bibinfo{volume}{3}}, \bibinfo{pages}{861} (\bibinfo{year}{2018}).
\newblock \urlprefix\url{https://doi.org/10.21105/joss.00861}.

\bibitem{stauffer1972universality}
\bibinfo{author}{Stauffer, D.}, \bibinfo{author}{Ferer, M.} \&
  \bibinfo{author}{Wortis, M.}
\newblock \bibinfo{title}{Universality of second-order phase transitions: the
  scale factor for the correlation length}.
\newblock \emph{\bibinfo{journal}{Phys. Rev. Lett.}}
  \textbf{\bibinfo{volume}{29}}, \bibinfo{pages}{345} (\bibinfo{year}{1972}).
\newblock \urlprefix\url{https://doi.org/10.1103/PhysRevLett.29.345}.

\bibitem{nikoghosyan2016universality}
\bibinfo{author}{Nikoghosyan, G.}, \bibinfo{author}{Nigmatullin, R.} \&
  \bibinfo{author}{Plenio, M.~B.}
\newblock \bibinfo{title}{Universality in the dynamics of second-order phase
  transitions}.
\newblock \emph{\bibinfo{journal}{Phys. Rev. Lett.}}
  \textbf{\bibinfo{volume}{116}}, \bibinfo{pages}{080601}
  (\bibinfo{year}{2016}).
\newblock \urlprefix\url{https://doi.org/10.1103/PhysRevLett.116.080601}.

\bibitem{gierlichs2008mutual}
\bibinfo{author}{Gierlichs, B.}, \bibinfo{author}{Batina, L.},
  \bibinfo{author}{Tuyls, P.} \& \bibinfo{author}{Preneel, B.}
\newblock \bibinfo{title}{Mutual information analysis}.
\newblock In \emph{\bibinfo{booktitle}{Cryptographic Hardware and Embedded
  Systems -- CHES 2008}}, \bibinfo{pages}{426--442}
  (\bibinfo{publisher}{Springer Berlin Heidelberg}, \bibinfo{address}{Berlin,
  Heidelberg}, \bibinfo{year}{2008}).
\newblock \urlprefix\url{https://doi.org/10.1007/978-3-540-85053-3_27}.

\bibitem{bau2020understanding}
\bibinfo{author}{Bau, D.} \emph{et~al.}
\newblock \bibinfo{title}{Understanding the role of individual units in a deep
  neural network}.
\newblock \emph{\bibinfo{journal}{Proc. Natl. Acad. Sci. U.S.A.}}
  \textbf{\bibinfo{volume}{117}}, \bibinfo{pages}{30071--30078}
  (\bibinfo{year}{2020}).
\newblock \urlprefix\url{https://doi.org/10.1073/pnas.190737511}.

\bibitem{simonyan2014very}
\bibinfo{author}{Simonyan, K.} \& \bibinfo{author}{Zisserman, A.}
\newblock \bibinfo{title}{Very deep convolutional networks for large-scale
  image recognition}.
\newblock In \bibinfo{editor}{Bengio, Y.} \& \bibinfo{editor}{LeCun, Y.} (eds.)
  \emph{\bibinfo{booktitle}{3rd International Conference on Learning
  Representations, {ICLR} 2015, San Diego, CA, USA, May 7-9, 2015, Conference
  Track Proceedings}} (\bibinfo{year}{2015}).
\newblock \urlprefix\url{https://doi.org/10.48550/arXiv.1409.1556}.

\bibitem{krizhevsky2014one}
\bibinfo{author}{Krizhevsky, A.}
\newblock \bibinfo{title}{One weird trick for parallelizing convolutional
  neural networks}.
\newblock \emph{\bibinfo{journal}{arXiv preprint arXiv:1404.5997}}
  (\bibinfo{year}{2014}).
\newblock \urlprefix\url{https://doi.org/10.48550/arXiv.1404.5997}.

\bibitem{tan2021efficientnetv2}
\bibinfo{author}{Tan, M.} \& \bibinfo{author}{Le, Q.}
\newblock \bibinfo{title}{Efficientnetv2: Smaller models and faster training}.
\newblock In \bibinfo{editor}{Meila, M.} \& \bibinfo{editor}{Zhang, T.} (eds.)
  \emph{\bibinfo{booktitle}{Proceedings of the 38th International Conference on
  Machine Learning}}, vol. \bibinfo{volume}{139} of
  \emph{\bibinfo{series}{Proceedings of Machine Learning Research}},
  \bibinfo{pages}{10096--10106} (\bibinfo{publisher}{PMLR},
  \bibinfo{year}{2021}).
\newblock \urlprefix\url{https://proceedings.mlr.press/v139/tan21a.html}.

\bibitem{Deng2009}
\bibinfo{author}{Deng, J.} \emph{et~al.}
\newblock \bibinfo{title}{Imagenet: A large-scale hierarchical image database}.
\newblock In \emph{\bibinfo{booktitle}{2009 IEEE Conference on Computer Vision
  and Pattern Recognition}}, \bibinfo{pages}{248--255}
  (\bibinfo{publisher}{IEEE}, \bibinfo{year}{2009}).
\newblock \urlprefix\url{https://doi.org/10.1109/CVPR.2009.5206848}.

\bibitem{torquato2002statistical}
\bibinfo{author}{Torquato, S.}
\newblock \bibinfo{title}{Statistical description of microstructures}.
\newblock \emph{\bibinfo{journal}{Ann. Rev. Mater. Res.}}
  \textbf{\bibinfo{volume}{32}}, \bibinfo{pages}{77--111}
  (\bibinfo{year}{2002}).
\newblock
  \urlprefix\url{https://doi.org/10.1146/annurev.matsci.32.110101.155324}.

\bibitem{niezgoda2013novel}
\bibinfo{author}{Niezgoda, S.~R.}, \bibinfo{author}{Kanjarla, A.~K.} \&
  \bibinfo{author}{Kalidindi, S.~R.}
\newblock \bibinfo{title}{Novel microstructure quantification framework for
  databasing, visualization, and analysis of microstructure data}.
\newblock \emph{\bibinfo{journal}{Integr. Mater. Manuf. Innov.}}
  \textbf{\bibinfo{volume}{2}}, \bibinfo{pages}{54--80} (\bibinfo{year}{2013}).
\newblock \urlprefix\url{https://doi.org/10.1186/2193-9772-2-3}.

\bibitem{bostanabad2018computational}
\bibinfo{author}{Bostanabad, R.} \emph{et~al.}
\newblock \bibinfo{title}{Computational microstructure characterization and
  reconstruction: Review of the state-of-the-art techniques}.
\newblock \emph{\bibinfo{journal}{Prog. Mater. Sci.}}
  \textbf{\bibinfo{volume}{95}}, \bibinfo{pages}{1--41} (\bibinfo{year}{2018}).
\newblock \urlprefix\url{https://doi.org/10.1016/j.pmatsci.2018.01.005}.

\bibitem{montes2021accelerating}
\bibinfo{author}{Montes~de Oca~Zapiain, D.}, \bibinfo{author}{Stewart, J.~A.}
  \& \bibinfo{author}{Dingreville, R.}
\newblock \bibinfo{title}{Accelerating phase-field-based microstructure
  evolution predictions via surrogate models trained by machine learning
  methods}.
\newblock \emph{\bibinfo{journal}{npj Comput. Mater.}}
  \textbf{\bibinfo{volume}{7}}, \bibinfo{pages}{1--11} (\bibinfo{year}{2021}).
\newblock \urlprefix\url{https://doi.org/10.1038/s41524-020-00471-8}.

\bibitem{aggarwal2001surprising}
\bibinfo{author}{Aggarwal, C.~C.}, \bibinfo{author}{Hinneburg, A.} \&
  \bibinfo{author}{Keim, D.~A.}
\newblock \bibinfo{title}{On the surprising behavior of distance metrics in
  high dimensional space}.
\newblock In \emph{\bibinfo{booktitle}{Database Theory -- International
  Conference on Database Theory 2001}}, \bibinfo{pages}{420--434}
  (\bibinfo{publisher}{Springer Berlin Heidelberg}, \bibinfo{address}{Berlin,
  Heidelberg}, \bibinfo{year}{2001}).
\newblock \urlprefix\url{https://doi.org/10.1007/3-540-44503-X_27}.

\bibitem{cogswell2011thermodynamic}
\bibinfo{author}{Cogswell, D.~A.} \& \bibinfo{author}{Carter, W.~C.}
\newblock \bibinfo{title}{Thermodynamic phase-field model for microstructure
  with multiple components and phases: The possibility of metastable phases}.
\newblock \emph{\bibinfo{journal}{Phys. Rev. E}} \textbf{\bibinfo{volume}{83}},
  \bibinfo{pages}{061602} (\bibinfo{year}{2011}).
\newblock \urlprefix\url{https://doi.org/10.1103/PhysRevE.83.061602}.

\bibitem{kim2012phase}
\bibinfo{author}{Kim, J.}
\newblock \bibinfo{title}{Phase-field models for multi-component fluid flows}.
\newblock \emph{\bibinfo{journal}{Commun. Comput. Phys.}}
  \textbf{\bibinfo{volume}{12}}, \bibinfo{pages}{613--661}
  (\bibinfo{year}{2012}).
\newblock \urlprefix\url{https://doi.org/10.4208/cicp.301110.040811a}.

\bibitem{weis1998morphological}
\bibinfo{author}{Weis, C.} \emph{et~al.}
\newblock \bibinfo{title}{Morphological and rheological detection of the phase
  inversion of {PMMA/PS} polymer blends}.
\newblock \emph{\bibinfo{journal}{Polym. Bull.}} \textbf{\bibinfo{volume}{40}},
  \bibinfo{pages}{235--241} (\bibinfo{year}{1998}).
\newblock \urlprefix\url{https://doi.org/10.1007/s002890050247}.

\bibitem{hedstrom2012phase}
\bibinfo{author}{Hedstr{\"o}m, P.}, \bibinfo{author}{Baghsheikhi, S.},
  \bibinfo{author}{Liu, P.} \& \bibinfo{author}{Odqvist, J.}
\newblock \bibinfo{title}{A phase-field and electron microscopy study of phase
  separation in {Fe}--{Cr} alloys}.
\newblock \emph{\bibinfo{journal}{Mater. Sci. Eng. A}}
  \textbf{\bibinfo{volume}{534}}, \bibinfo{pages}{552--556}
  (\bibinfo{year}{2012}).
\newblock \urlprefix\url{https://doi.org/10.1016/j.msea.2011.12.007}.

\bibitem{Gal2017}
\bibinfo{author}{Gal, Y.}, \bibinfo{author}{Hron, J.} \&
  \bibinfo{author}{Kendall, A.}
\newblock \bibinfo{title}{Concrete dropout}.
\newblock In \bibinfo{editor}{Guyon, I.} \emph{et~al.} (eds.)
  \emph{\bibinfo{booktitle}{Advances in Neural Information Processing
  Systems}}, vol.~\bibinfo{volume}{30} (\bibinfo{publisher}{Curran Associates,
  Inc.}, \bibinfo{year}{2017}).
\newblock
  \urlprefix\url{https://proceedings.neurips.cc/paper/2017/file/84ddfb34126fc3a48ee38d7044e87276-Paper.pdf}.

\bibitem{chen2002phase}
\bibinfo{author}{Chen, L.-Q.}
\newblock \bibinfo{title}{Phase-field models for microstructure evolution}.
\newblock \emph{\bibinfo{journal}{Ann. Rev. Mater. Res.}}
  \textbf{\bibinfo{volume}{32}}, \bibinfo{pages}{113--140}
  (\bibinfo{year}{2002}).
\newblock
  \urlprefix\url{https://doi.org/10.1146/annurev.matsci.32.112001.132041}.

\bibitem{dingreville2020benchmark}
\bibinfo{author}{Dingreville, R.}, \bibinfo{author}{Stewart, J.~A.},
  \bibinfo{author}{Chen, E.~Y.} \& \bibinfo{author}{Monti, J.~M.}
\newblock \bibinfo{title}{Benchmark problems for the mesoscale multiphysics
  phase field simulator ({MEMPHIS}).}
\newblock \bibinfo{type}{Tech. Rep.} \bibinfo{number}{SAND2020-12852},
  \bibinfo{institution}{Sandia National Laboratories (SNL-NM)},
  \bibinfo{address}{Albuquerque, NM, USA} (\bibinfo{year}{2020}).
\newblock \urlprefix\url{https://doi.org/10.2172/1615889}.

\bibitem{He2016a}
\bibinfo{author}{He, K.}, \bibinfo{author}{Zhang, X.}, \bibinfo{author}{Ren,
  S.} \& \bibinfo{author}{Sun, J.}
\newblock \bibinfo{title}{Deep residual learning for image recognition}.
\newblock In \emph{\bibinfo{booktitle}{2016 {IEEE} Conference on Computer
  Vision and Pattern Recognition, {CVPR} 2016, Las Vegas, NV, USA, June 27-30,
  2016}}, \bibinfo{pages}{770--778} (\bibinfo{publisher}{{IEEE} Computer
  Society}, \bibinfo{year}{2016}).
\newblock \urlprefix\url{https://doi.org/10.1109/CVPR.2016.90}.

\bibitem{Bengio1994}
\bibinfo{author}{Bengio, Y.}, \bibinfo{author}{Simard, P.} \&
  \bibinfo{author}{Frasconi, P.}
\newblock \bibinfo{title}{Learning long-term dependencies with gradient descent
  is difficult}.
\newblock \emph{\bibinfo{journal}{{IEEE} Trans. Neural Netw.}}
  \textbf{\bibinfo{volume}{5}}, \bibinfo{pages}{157--166}
  (\bibinfo{year}{1994}).
\newblock \urlprefix\url{https://doi.org/10.1109/72.279181}.

\bibitem{dAscoli2020}
\bibinfo{author}{d'Ascoli, S.}, \bibinfo{author}{Sagun, L.} \&
  \bibinfo{author}{Biroli, G.}
\newblock \bibinfo{title}{Triple descent and the two kinds of overfitting:
  where {\&} why do they appear?}
\newblock In \bibinfo{editor}{Larochelle, H.}, \bibinfo{editor}{Ranzato, M.},
  \bibinfo{editor}{Hadsell, R.}, \bibinfo{editor}{Balcan, M.} \&
  \bibinfo{editor}{Lin, H.} (eds.) \emph{\bibinfo{booktitle}{Advances in Neural
  Information Processing Systems 33: Annual Conference on Neural Information
  Processing Systems 2020, NeurIPS 2020, December 6-12, 2020, virtual}}
  (\bibinfo{year}{2020}).
\newblock
  \urlprefix\url{https://proceedings.neurips.cc/paper/2020/hash/1fd09c5f59a8ff35d499c0ee25a1d47e-Abstract.html}.

\end{thebibliography}
%%%%%%%%%%%%%%%%%%%%%%%%%%%%%%%%%%%%%%%%%%%%%%%%%%%%%%%%%%%%%%%%%
%%%%%%%%%%%%%%%%%%%%%%%%%%%%%%%%%%%%%%%%%%%%%%%%%%%%%%%%%%%%%%%%%
%%
%% ACKNOWLEDGMENTS
%%
%%%%%%%%%%%%%%%%%%%%%%%%%%%%%%%%%%%%%%%%%%%%%%%%%%%%%%%%%%%%%%%%%
%%%%%%%%%%%%%%%%%%%%%%%%%%%%%%%%%%%%%%%%%%%%%%%%%%%%%%%%%%%%%%%%%
\section*{ACKNOWLEDGMENTS}
\label{sec:Acknowledgments}
{
The authors would like to thank M D'Elia from Sandia National Laboratories for comments and review of this work.
MA acknowledges funding from USC-ISI and Sandia National Laboratories. KB also acknowledges funding from Sandia National Laboratories. 
RD acknowledges funding under the \textit{Beyond}Fingerprinting Sandia Grand Challenge Laboratory Directed Research and Development (GC LDRD) program.
The phase-field framework is supported by the Center for Integrated Nanotechnologies (CINT), an Office of Science user facility operated for the U.S. Department of Energy. 
Sandia National Laboratories is a multi-mission laboratory managed and operated by National Technology and Engineering Solutions of Sandia, LLC., a wholly owned subsidiary of Honeywell International, Inc., for the U.S.\ Department of Energy National Nuclear Security Administration under contract DE-NA0003525. This paper describes objective technical results and analysis. Any subjective views or opinions that might be expressed in the paper do not necessarily represent the views of the U.S.\ Department of Energy or the United States Government.
}
%%%%%%%%%%%%%%%%%%%%%%%%%%%%%%%%%%%%%%%%%%%%%%%%%%%%%%%%%%%%%%%%%
%%%%%%%%%%%%%%%%%%%%%%%%%%%%%%%%%%%%%%%%%%%%%%%%%%%%%%%%%%%%%%%%%
%%
%% CONTRIBUTION
%%
%%%%%%%%%%%%%%%%%%%%%%%%%%%%%%%%%%%%%%%%%%%%%%%%%%%%%%%%%%%%%%%%%
%%%%%%%%%%%%%%%%%%%%%%%%%%%%%%%%%%%%%%%%%%%%%%%%%%%%%%%%%%%%%%%%%

\section*{AUTHOR CONTRIBUTIONS}
\label{sec:Contribution}
{
MA, KB, RD, JS, GVS, and AG conceived the research, contributed to the investigation, and wrote the manuscript. MA and KB developed the methodology and visualization. Phase-field simulations were performed at CINT. RD, GVS, and AG acquired funding, administered the project, and supervised the research.
}
%%%%%%%%%%%%%%%%%%%%%%%%%%%%%%%%%%%%%%%%%%%%%%%%%%%%%%%%%%%%%%%%%:~

\section*{COMPETING INTERESTS}

    The authors declare no competing interests.

%%%%%%%%%%%%%%%%%%%%%%%%%%%%%%%%%%%%%%%%%%%%%%%%%%%%%%%%%%%%%%%%%:

\clearpage

    

\section*{Supplementary material}

# Supplementary Information for:
# Inferring topological transitions in pattern-forming processes with self-supervised learning

**Marcin Abram**[a,b,*], **Keith Burghardt**[b,†], **Greg Ver Steeg**[b], **Aram Galstyan**[b], **and Rémi Dingreville**[c,‡]

[a]Department of Physics and Astronomy, University of Southern California, Los Angeles, CA 90089, USA
[b]Information Sciences Institute, University of Southern California, Marina del Rey, CA 90292, USA
[c]Center for Integrated Nanotechnologies, Nanostructure Physics Department, Sandia National Laboratories, Albuquerque, NM 87185, USA
[*]mjarbam@usc.edu
[†]keithab@isi.edu
[‡]rdingre@sandia.gov

## ABSTRACT

This supplementary information contains details on: (i) the ambiguity of identifying transitions regardless of the latent projection technique used (Supplementary Note 1), (ii) the robustness of the approach presented in the main manuscript when using other latent projection techniques and neural network approaches (Supplementary Note 2), (iii) examples of raw data used in this work (Supplementary Note 3), (iv) details on the training data used for the pre-trained ResNet-50 v2 (Supplementary Note 4), and (v) details on the training performance of the ResNet-50 v2 model used in the main manuscript.

## Contents

## Supplementary Note 1: Ambiguity of identifying transitions using different types of latent projections

The ResNet-50 v2 model used for projecting our original input was trained on the ImageNet dataset (see Supplementary Note 4 for further details). This model was not fine-tuned any further. This decision was by design to highlight how robust our method is to the choice of the projection method. While it might be somehow surprising results that we could adopt a model pre-trained on natural images to another scientific domain without any fine-tuning, it can be easily explained. The convolutional neural network (CNN) learns to detect and combine different patterns. This approach works well to detect simple geometric shapes as it is the case with the data we deal with (see Supplementary Note 3 for instance).

To further show that the different projection methods lead to similar results, we trained (from scratch, using an unsupervised method described based on a discriminative unsupervised feature learning[1]) two networks: one simple CNN and one more sophisticated, based on the ResNet-34 architecture. We also took the pre-trained ResNet-50 v2 and fine-tuned it on our data. We visualize the latent space using three techniques, (i) by showing the first two most meaningful principal components from principal component analysis (PCA), (ii) by applying uniform manifold approximation and projection (UMAP) technique, (iii) by constructing the t-distributed stochastic neighbor embedding (t-SNE), see **Suppl. Fig. 1**.

All models tested produced qualitatively similar embedding as the ones presented in the main manuscript. It is remarkable, that pre-trained ResNet-50 v2, trained on different domain (ImageNet) produces as good projection as network that were trained to learn the input features from scratch. The reason might be related to the fact that the inputs are relatively simple. Thus, network trained to differentiate between different natural objects (like animals, plants, everyday objects) can also differentiate between simple geometric patterns (vertical *vs*. horizontal domains, domains of different width, domains of different size and shape, *etc*.).

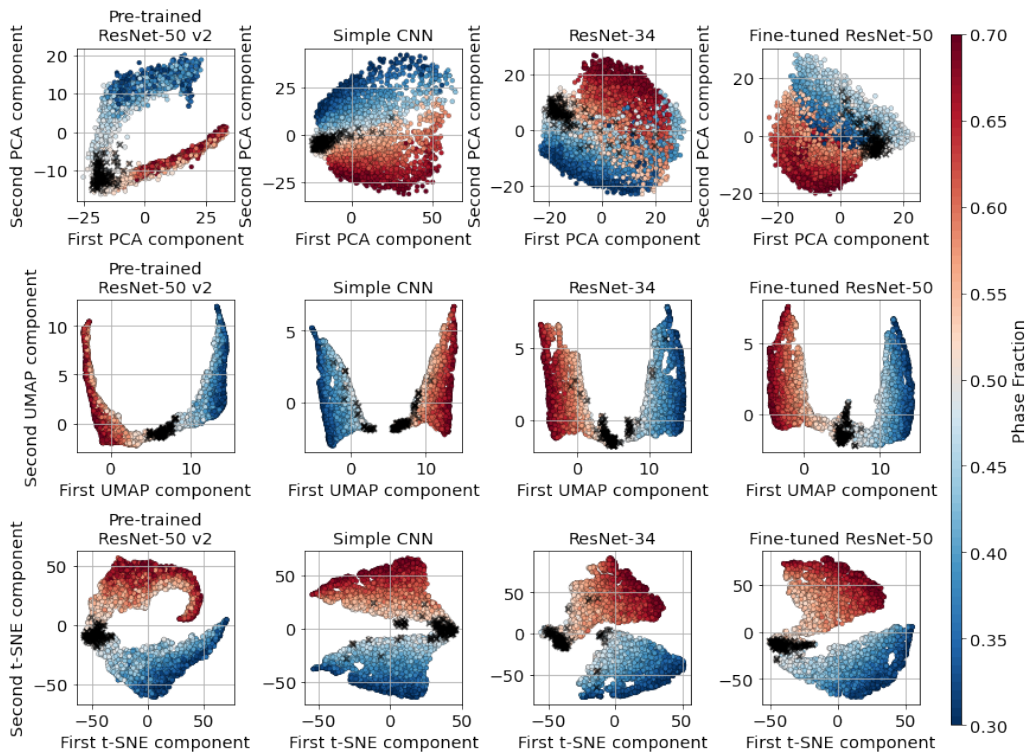

**Supplementary Figure 1. Visualization of the latent space for spinodal decomposition data obtained using different projection methods.** In the first column we show the latent space as produced by a pre-trained ResNet-50 v2 (no fine-tuning). In the second and the third columns, the latent space constructed by a simple CNN and ResNet-34 model, respectively, trained to discriminate inputs representing different experiments. The fourth column represents the latent space as constructed after fine-tuning the pre-trained ResNet-50 v2. We visualize the latent space using three techniques, (i) by showing the first two most meaningful PCA components, (ii) by applying uniform manifold approximation and projection (UMAP) technique, (iii) by constructing the t-distributed stochastic neighbor embedding (t-SNE). Note, that both UMAP and t-SNE are stochastic methods, while PCA is deterministic. UMAP captures both global and local structure of the data, while t-SNE only the local structure (thus, see the disconnected graph for ResNet-34 in the last row).

## Supplementary Note 2: Robustness of solving the inverse problem

To show that the pre-trained ResNet-50 v2 generalizes well to different scientific domains, we solved the inverse problem described in the main manuscript using different projection methods described in Supplementary Note 1. We present these results in **Suppl. Fig.** 2.

We contrast the results presented in the main manuscript reproduced in **Suppl. Fig.** 2**a** with those obtained using a fine-tuned model depicted in **Suppl. Fig.** 2**d**. We observe that we get comparable results to identify the transition in pattern formation as a function of the input process parameters. We also observe that the change in the slope of the measured sensitivity score is more evident in the latter case. Thus, while fine-tuning might help us to get clearer results, at least in this case it does not qualitatively change the observed outcome. Similarly, by contrasting **Suppl. Fig.** 2**a** with **Suppl. Figs.** 2**b**–**c**, we can compare the difference between a general projection model and a dedicated model trained directly on the data of interests. Also in this case, we do not see any significant differences. This might be related to the fact that in the considered case the input consists of relatively simple geometric structures. Perhaps, for more challenging problems, the difference between a general and dedicated methods would be more significant. While this is an important consideration to keep in mind, in general, the fact that a general pre-trained model produces a sufficient projection provides credence to our general approach. It means that our framework is robust to the choice of the projection method. It lowers the computational requirements of the method (as it is not necessary to train its own projection model). Finally, it also lowers the entry barrier for adoption to other scientific domains, since different versions of pre-trained models are readily available, *e.g.*, on PyTorch or TensorFLow Hubs.

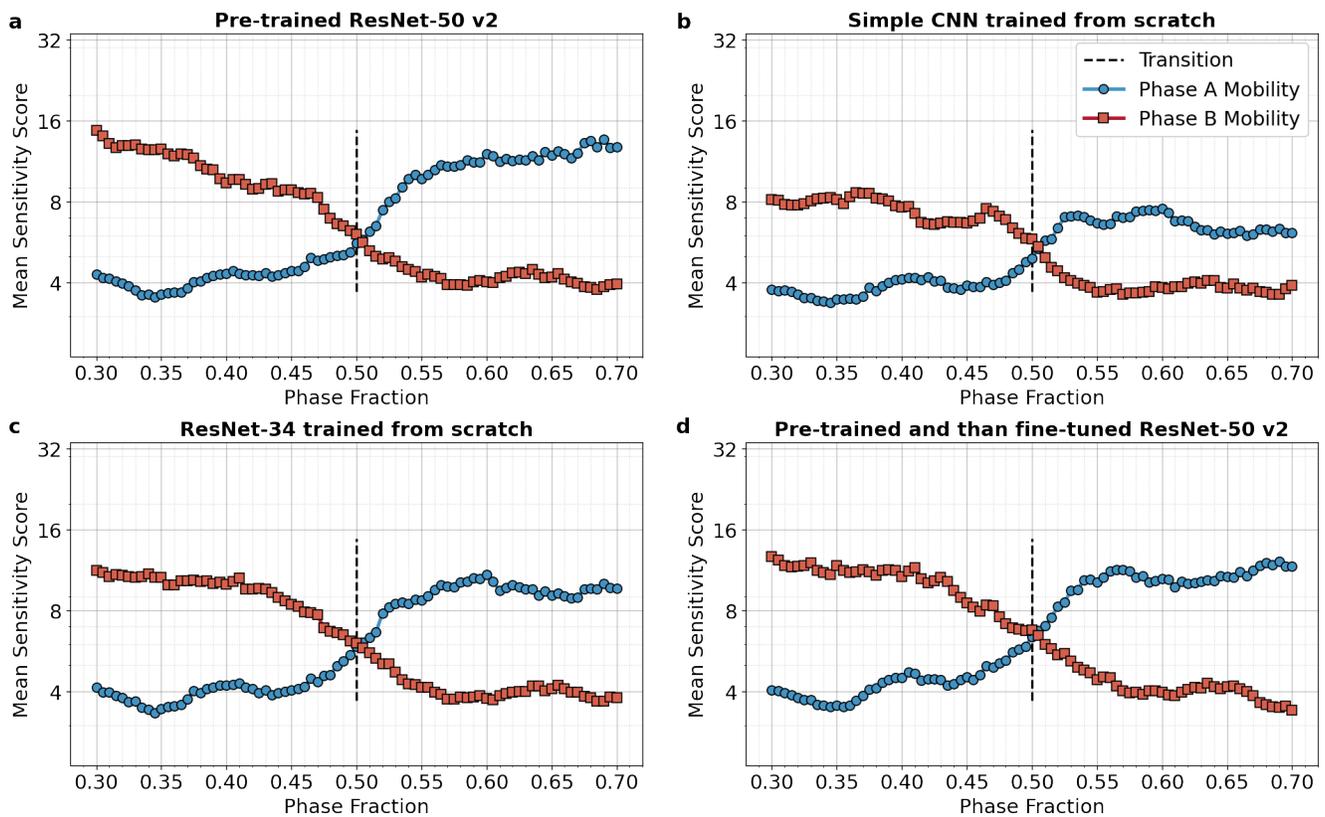

**Supplementary Figure 2. Transitions predicted for different projection methods. a** Using pre-trained on ImageNet (and not fine-tuned) ResNet-50 v2. **b** Using a simple CNN model, trained from scratch using a discriminative unsupervised feature learning method[1]. **c** Using a more sophisticated model, ResNet-34 (a smaller version of ResNet-50), trained from scratch using the same method as the simple CNN. **d** Using a ResNet-50 v2 model, pre-trained on ImageNet, fine-tuned on the data of interests.

# Supplementary Note 3: Examples of raw simulation data

In **Suppl. Fig. 3a**, we show raw simulation data for the two-dimensional spatio-temporal evolution of the microstructural patterns for the spinodal decomposition problem. The simulation is initiated with a random seed in $T = 0$. As time increases, we observe separation of the phases and the formation of specific microstructural patterns. In **Suppl. Fig. 3b**, we show raw data for the formation of microstructural patterns for the phase-field model pertaining to the physical vapor deposition of thin films. As time increases, the thickness of the solid thin film increases. Only the microstructural patterns forming in the solid phase of film domain are used for this work.

The final state of the simulation depends both on the initial conditions (initial distribution of the phases determined by a random seed) and the value of the process parameters. While the specific local details of particular microstructural patterns are very sensitive to the initial conditions, other characteristics of those microstructures (*e.g.*, the shape and size of the phase domains, the orientation and thickness of the subdomains) is determined by the value of the process parameters, independent of what the initial microstructure was at time $t = 0$.

When solving the inverse problem, we focus on the prediction of the latter: predicting the process parameters not the initial (random) conditions. In turn, this type of predictions determines our choice of machine-learning models. To embed the raw data, we used a version of a convolutional neural network (ResNet-v2) that is equivariant to translations such that our approach not sensitive to specific positions of the observed patterns, but rather on other characteristics like size, orientation or shape of the subdomains.

It is also worth mentioning that since the input to our machine-learning model is constructed from the latter stages of the simulation, the relative impact of the initial conditions on the character of the formed microstructural patterns should be minimal (*e.g.*, the orientation of the subdomains is mostly determined by the values of the input process parameters, not the random initialization of the simulation).

The initial dimensionality of the raw data is $512 \times 512 = 262,144$ and $256 \times 256 = 65,536$ for spinodal decomposition and the physical vapor deposition simulations, respectively. While possible, working directly on such a large input set would be extremely inefficient. Additionally, not every detail included in the raw input is relevant in the context of our task of predicting transition in pattern formation. While size, shape and orientation of the microstructures are related to the values of some process parameters (like phase fraction, phase mobility, deposition rate, deposition angle, *etc.*), other characteristics of the input, like specific state of the initial microstructures, might be only related to the specific random choice of the initial state (*cf.* **Suppl. Fig. 3**). While it is possible to manually construct a smaller set of features that can be used to characterize the type of observed microstructure (*e.g.*, orientation of the domains in the PVD case would be one of possible choice), there might be several drawbacks of such an approach, including (i) the need to construct a separate set of features for each physical process, (ii) the lack of knowledge on which features are the most informative and which can be ignored, and finally (iii) the possible introduction of human bias when selecting features by hand. Instead, we employ neural network trained to extract important features from visual data, such as a ResNet-50 v2 trained on ImageNet.

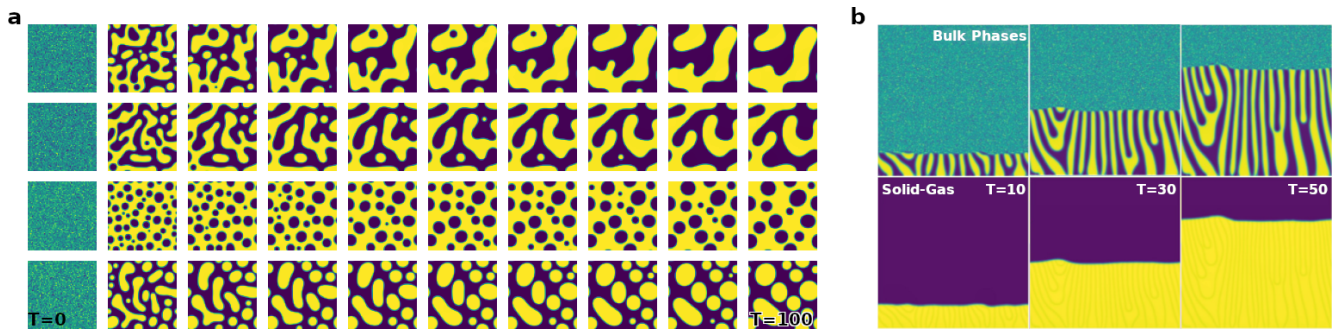

**Supplementary Figure 3. Raw data examples**. **a** Raw simulation data for the two-dimensional spatio-temporal evolution of the microstructural patterns for the spinodal decomposition problem. **b** Raw data the two-dimensional spatio-temporal evolution of the microstructural patterns during physical vapor deposition of thin films.

## Supplementary Note 4: Training data for the ResNet-50 v2 model

ResNet is a type of a deep-convolutional neural network (CNN), that incorporates skip-connections to reduce the vanishing gradient problem during the training. It is based on repeated residual blocks, that allow to model a deviation of the signal from a typical value instead of the absolute change of the signal (this approach drastically increases stability of the training). The version of the ResNet used in the paper has a depth of 50 layers. We used an updated version of that architecture (v2)[2], that has a re-arranged order of the layers in the residual block, in comparison to the initial implementation of that model[3].

The version of ResNet-50 v2 that we used in the main manuscript was trained on the ImageNet dataset[4]. We show an example of the training data from the ImageNet dataset in Suppl. Fig. 4**a**. These examples depict 1000 classes such as different types of animals (goldfish, shark, hen, tick, koala, *etc*.), everyday objects (dumbbell, desk, golf ball, iPad, French horn, *etc*.), and various building structures (barbershop, bakery, greenhouse, prison, restaurant, *etc*.). While most of the training examples represent colorful (3-channel) images, there is also a small proportion (about ∼2%) of monochromatic images, as presented in **Suppl. Fig. 4b**.

The visual examples used to train the ResNet model are drastically different from the scientific data analyzed in this paper (*cf*. **Suppl. Fig. 4c–d**). However, as demonstrated in the main manuscript, generalization from one domain of visual input to another is possible. The model trained on the ImageNet dataset learnt to extract a hierarchy of features, from simple geometric shapes (vertical, horizontal, or skew lines, shapes of different sizes, lines of different curvatures, *etc*.) to more complex, composed patterns. A model that is trained on diverse set of natural images can effectively extract patterns from (visually) simpler domain, such as the results of the spatio-temporal evolution of the microstructural patters (contrast **Suppl. Figs. 4a–b** with 4**c–d**).

Another technical challenge, that we had to address in our work was that while the ResNet model used was trained on colorful images (see **Suppl. Fig. 4a**), the phase-field simulation data represent a monochromatic (one-channel) data (see **Suppl. Fig. 4c–d**). To make the phase-field simulation data input compatible with our model, we copied each image three times (creating 3 identical pseudo-color channels). While it created some redundancy (the network might detect "red", "green" and "blue" edges of the same orientation) it did not affect the ability to produce a consistent embedding for the collection of our data (as evident in our comparison, and as discussed in the next section).

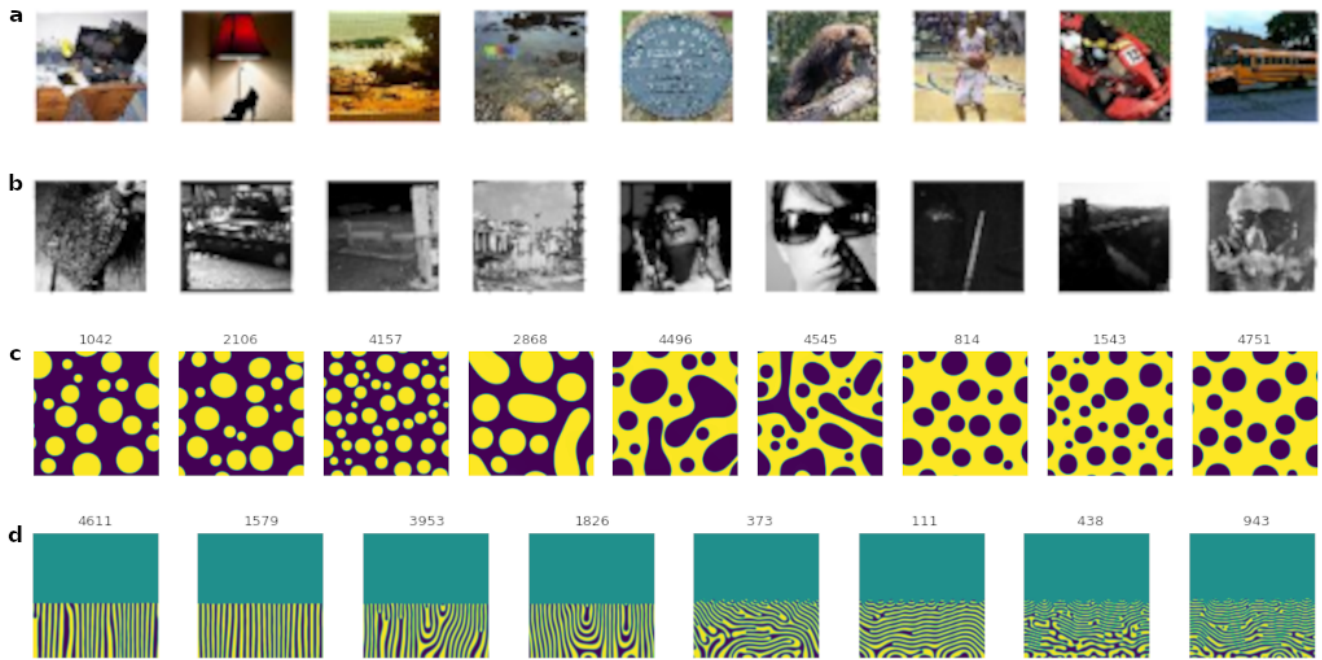

**Supplementary Figure 4. Training Data Examples. a** Training example from the ImageNet dataset, used to train ResNet-50 v2 model. **b** Selection of monochromatic training examples from ImageNet dataset. **c** Evolved state of the spinodal decomposition process (after a fixed number of steps of time-evolution). **d** Evolved state of the physical vapor deposition process (fixed with respect to the width of the solid section).

## Supplementary Note 5: Training performance of ResNet-50 v2 model

When solving the inverse problem, we used a dense, forward-fed network with two hidden layers and 512 and 1024 neurons, respectively. Depending on the dimension of the latent space (2048 for ResNet-50, 512 for custom CNN and ResNet-34), the number of parameters of the neural network ranged from 790,018 to 1,576,450. This is however, a typical situation for neural networks to have more parameters than data samples. As evident in the S. d'Ascoli, L. Sagun, and G. Biroli, "Triple descent and the two kinds of overfitting: Where & why do they appear?"[5], the largest overfitting happens when the number of parameters is comparable with the number of training examples. If the number of parameters of the network is either much smaller or much larger than the number of training examples, the generalization error should be relatively small (if the model is trained properly).

To minimize any potential overfitting, we followed the best practices during the training. Namely, we used a validation set to track the generalization error, we used dropout to combat the overfitting, we used data augmentation (we took advantage of the periodic boundaries conditions, and we prepared several cropped versions of the input from each sample), and when training the model we used an early stop. We visualize the training and validation error for each projection method in **Suppl. Fig.** 5 (for brevity of that supplement, we show only results obtained for spinodal decomposition data).

The important realization here is that when solving the inverse problem, we note two types of errors, epistemic and aleatoric. While we can reduce the epistemic error, we cannot eliminate the aleatoric one. Therefore it is expected that for some process parameters, in some regions, we will observe a large error. This is especially visible when analyzing the spinodal decomposition case. For example, when the phase fraction is $< 0.5$, the character of the output is determined by the value of phase B and it is invariant to the value of phase A (*cf.* Fig. 4c-d in the main manuscript). Thus, even if we had an ideal model and a perfect training procedure, the best we can hope is to have zero error, $\Delta_B$, when predicting phase B and an error of $\Delta_A = \int_0^1 (x-0.5)^2\, dx = 1/12$ when predicting values of phase A (assuming that our values are normalized to [0, 1] and we have a uniform distribution of samples – in such a case to minimize the mean squared error a perfect learner would output mean values, 0.5). Therefore, even a perfect learner cannot reduce the error below a certain level. Hence, we expect to see a large error for some target parameters in the regions where those parameters have no influence on the character of the final output. We have illustrated this point in Table 1 in the main manuscript. To further analyze that our network learned to correctly solve the inverse problem (as well as possible), we show the mean squared error on normalized target approaches after a few epochs the expected minimal value (see the **Suppl. Figs.** 5**a–d**), where the theoretical minimum is

$$\bar{\Delta} = \frac{\Delta_A + \Delta_B}{2} = \frac{1}{2}\left(\frac{1}{12} + 0\right) \approx 0.0416666\ldots \tag{S.1}$$

We see that in each case the validation set approaches the limits of the irreducible/intrinsic error, indicating that we were able to learn what there is to learn (with minimal amount of overfitting and relatively good generalization capability). While the results can be further improved by careful fine-tuning of the training hyper-parameters, such optimization choices are not crucial. What really matters in our approach is the relative comparison of the errors for different process parameters. Thus, even if there is another network that has a lower generalizability error, it does not make the present results invalid or incorrect. As long as the predictions are consistent, we can compare the change of the prediction errors and detect where the transitions are. This is actually similar to the "learning by confusion" schema by van Nieuwenburg et al.[6], where the results are robust to the size of the network, since what matter is the relative maximum of the accuracy, not the absolute value.

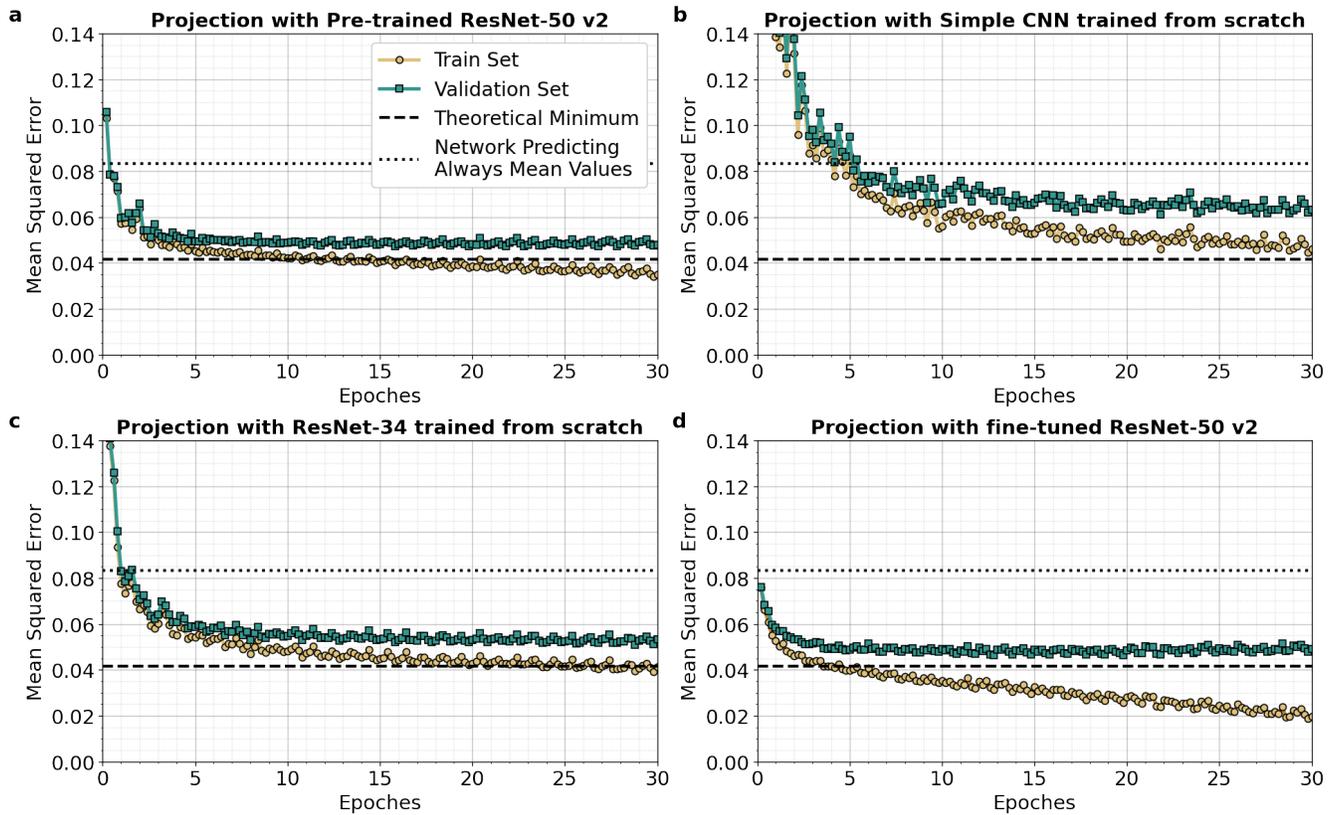

**Supplementary Figure 5. Training logs for different projection methods. a** With data embedded using ResNet-50 v2 pre-trained on ImageNet. **b** With data embedded using a simple CNN model, trained from scratch using a discriminative unsupervised feature learning method[1]. **c** With data embedded using ResNet-34, trained from scratch using the same method as the simple CNN. **d** With data embedded using the pre-trained and fine-tuned ResNet-50 v2.

## Supplementary References

1. Dosovitskiy, A., Fischer, P., Springenberg, J. T., Riedmiller, M. A. & Brox, T. Discriminative unsupervised feature learning with exemplar convolutional neural networks. *IEEE Trans. Pattern Anal. Mach. Intell.* **38**, 1734–1747 (2016). URL https://doi.org/10.1109/TPAMI.2015.2496141.

2. He, K., Zhang, X., Ren, S. & Sun, J. Identity mappings in deep residual networks. In *Computer Vision – ECCV 2016*, 630–645 (Springer International Publishing, 2016). URL https://doi.org/10.1007/978-3-319-46493-0_38.

3. He, K., Zhang, X., Ren, S. & Sun, J. Deep residual learning for image recognition. In *2016 IEEE Conference on Computer Vision and Pattern Recognition, CVPR 2016, Las Vegas, NV, USA, June 27-30, 2016*, 770–778 (IEEE Computer Society, 2016). URL https://doi.org/10.1109/CVPR.2016.90.

4. Deng, J. *et al.* Imagenet: A large-scale hierarchical image database. In *2009 IEEE Conference on Computer Vision and Pattern Recognition*, 248–255 (IEEE, 2009). URL https://doi.org/10.1109/CVPR.2009.5206848.

5. d'Ascoli, S., Sagun, L. & Biroli, G. Triple descent and the two kinds of overfitting: where & why do they appear? In Larochelle, H., Ranzato, M., Hadsell, R., Balcan, M. & Lin, H. (eds.) *Advances in Neural Information Processing Systems 33: Annual Conference on Neural Information Processing Systems 2020, NeurIPS 2020, December 6-12, 2020, virtual* (2020). URL https://proceedings.neurips.cc/paper/2020/hash/1fd09c5f59a8ff35d499c0ee25a1d47e-Abstract.html.

6. van Nieuwenburg, E. P. L., Liu, Y.-H. & Huber, S. D. Learning phase transitions by confusion. *Nat. Phys.* **13**, 435–439 (2017). URL https://doi.org/10.1038/nphys4037.



\end{document}